\documentstyle[preprint,amssymb,aps,tighten,epsf,floats]{revtex}
\newcommand{\triton}{$^3\mbox{H}$}

\newcommand{\nn}{\nonumber\\}
\newcommand{\ggama}{{\scriptstyle\Gamma}}
\newcommand{\hggama}{{\scriptstyle\Gamma}}
\newcommand{\la}{\langle}
\newcommand{\ra}{\rangle}

\newcommand{\bPsi}{\bar{\Psi}}
\newcommand{\bpsi}{\bar{\psi}}
\newcommand{\Psin}{\Psi^{\mbox{\scriptsize in}}}

\newcommand{\bPsout}{\bar{\Psi}^{\mbox{\scriptsize out}}}
\newcommand{\iin}{\mbox{\normalsize in}}
\newcommand{\oout}{\mbox{\normalsize out}}
\newcommand{\ben}{\begin{displaymath}}
\newcommand{\een}{\end{displaymath}}
\newcommand{\be}{\begin{equation}}
\newcommand{\ee}{\end{equation}}
\newcommand{\bea}{\begin{eqnarray}}
\newcommand{\eea}{\end{eqnarray}}

\newcommand{\he}{\hat{e}}

\newcommand{\hw}{{w}}

\newcommand{\piN}{$\pi N$}
\newcommand{\bphi}{\bar{\phi}}

\newcommand{\NN}{$N\!N$}
\newcommand{\NNpiNN}{$N\!N\!-\pi N\!N$}
\newcommand{\piNN}{$\pi N\!N$}

\newcommand{\NNN}{$N\!N\!N$}

\newcommand{\TT}{{\cal T}}
\newcommand{\UU}{{\cal U}}
\newcommand{\WW}{{\cal W}}
\newcommand{\tUU}{\tilde{\cal U}}
\newcommand{\tU}{\tilde{U}}

\newcommand{\tT}{\tilde{T}}
\newcommand{\mbf}[1]{\mbox{\boldmath {#1}}}

\newcommand{\bc}{\begin{center}}
\newcommand{\ec}{\end{center}}

\newcommand{\eqn}[1]{\label{#1}}
\newcommand{\eq}[1]{Eq.~(\ref{#1})}
\newcommand{\eqs}[1]{Eqs.~(\ref{#1})}
\newcommand{\fign}[1]{\label{#1}}
\newcommand{\fig}[1]{Fig.~\ref{#1}}

\begin{document}
\draft
\title{\bf Gauging of equations method.\\
I. Electromagnetic currents of three distinguishable particles}
\author{A. N. Kvinikhidze\footnote{On leave from Mathematical Institute of
Georgian Academy of Sciences, Tbilisi, Georgia.} and B. Blankleider}
\address{Department of Physics, The Flinders University of South Australia,
Bedford Park, SA 5042, Australia}
\date{\today}
\maketitle
\begin{abstract}
We present a general method for incorporating an external electromagnetic field
into descriptions of few-body systems whose strong interactions are described by
integral equations. In particular, we address the case where the integral
equations are those of quantum field theory and effectively sum up an infinite
number of Feynman diagrams. The method involves the idea of gauging the integral
equations themselves, and results in electromagnetic amplitudes where an
external photon is effectively coupled to every part of every strong interaction
diagram in the model.  Current conservation is therefore implemented in the
way prescribed by quantum field theory. We apply our gauging procedure to the
four-dimensional integral equations describing a system of three distinguishable
relativistic particles. In this way we obtain the expressions needed to
calculate all possible electromagnetic processes of the three-body system. An
interesting aspect of our results is the natural appearance of a subtraction
term needed to avoid the overcounting of diagrams.
\end{abstract}


\section{Introduction}
With the advent of new experimental facilities like the Thomas Jefferson
National Accelerator Facility (TJAF), the Electron Stretcher Accelerator (ELSA),
and the Mainz Microtron (MUMI), there is currently great interest in the use of
photons and electrons to probe the structure of hadronic systems. In practice,
this means using the electromagnetic probes to induce a variety of reactions
amongst the hadrons. Here we shall be concerned with those reactions where
effectively only one external photon is involved.  This includes not only
photoabsorption and photoproduction reactions, but also electron scattering and
electroproduction when calculated to lowest order in the electromagnetic
interaction. On the theoretical side such reactions are described by $\la
J^\mu\ra$, the matrix element of the electromagnetic current operator, and it
must ultimately be the goal of the theorist to construct models where $\la
J^\mu\ra$ describes the experimental data as accurately as possible.

An essential constraint on $\la J^\mu\ra$ is that it must obey current
conservation, expressed by the continuity equation $\partial _\mu \la J^\mu \ra
=0$. Current conservation is a consequence of charge conservation and is
therefore a fundamental property of any theory. On the other hand, because one
uses models to approximate the full theory, current conservation is not
guaranteed {\em a priori}. For this reason, much effort has been devoted to the
question of how to impose current conservation within a particular model of
strong interactions. In a seminal paper, Gross and Riska (GR) \cite{GR} have
shown how to construct the conserved current of two nucleons described by the
Bethe-Salpeter (BS) equation.  By analysing specific meson-exchange diagrams of
the BS kernel, they showed that current conservation is achieved when the
two-nucleon interaction current is constructed by attaching photons to all
possible places in the BS kernel. Although indispensable for the case of a
two-body system, this result does not apply to systems consisting of three or
more particles. There is also no straightforward way to use this result to
construct the electromagnetic current of three particles even in the case where
only two-body strong interactions are present.

Here we would therefore like to present a different method for constructing
conserved currents that is applicable to any number of particles whose strong
interaction processes are described by relativistic four-dimensional integral
equations. This method involves a direct gauging of the equations themselves in
the sense that a vector index $\mu$ is added to all terms of the equations in
such a way that a linear equation in $\mu$-labelled quantities results.
Proceeding in this way, we obtain integral equations for the gauged quantities
of interest (e.g. the gauged Green function $G^\mu$ or the gauged scattering
$t$-matrix $T^\mu$) expressed in terms of other gauged quantities that are known
or that can be easily constructed (typically the gauged one-particle propagator
$d^\mu$ and the gauged two-particle potential $v^\mu$). This approach, which we
shall simply refer to as {\em gauging the equations}, results in the external
photon being effectively coupled everywhere in the strong interaction model, so
that current conservation is guaranteed.

Using this method for the case of two nucleons, we obtain that the hadronic
current is a sum of matrix elements of the gauged nucleon propagators and the
gauged BS kernel. Thus our result, in this case, coincides with the one of
GR. However, our method is also easily applied to other systems. As the simplest
strongly interacting system going beyond the two-body problem is that of three
particles, we choose to illustrate the general nature of our gauging method by
applying it to the relativistic three-body problem whose strong interactions are
described by standard four-dimensional integral equations. That is, we consider
systems like that of three nucleons or three quarks whose strong interaction
processes do not involve coupling to the two-body sector. To keep the
presentation as simple as possible we restrict the discussion to the case of
three distinguishable particles. The gauging of three identical particles
involves additional considerations of particle exchange symmetry and is the
subject of the accompanying paper \cite{II}.

Although four-dimensional three-body equations have already been pursued
numerically by Rupp and Tjon \cite{RT}, there has not been a generalisation of
the Gross-Riska result to the case of three particles. As a result, there is
presently no rigorous derivation of the conserved current for a relativistic
three-body systems. This paper is therefore devoted to showing how the gauging
of relativistic three-body equations leads to gauge invariant expressions for
the various electromagnetic transition currents of a three-particle system. It
is a feature of our approach that the gauge invariance of our expressions is not
imposed in an {\em ad hoc} fashion, but rather according to the way prescribed
by field theory, namely by coupling photons to all possible places in the strong
interaction model.

To show the flexibility of the gauging method, we apply it to two different
relativistic three-body equations. The first of these is the integral equation
for the $3\rightarrow 3$ Green function whose kernel is defined in terms of
two-body potentials. This leads to three-body electromagnetic currents which are
expressed in terms of two-body potentials $v$ and gauged two-body potentials
$v^\mu$. In the second approach, we gauge the four-dimensional version of the
Alt-Grassberger-Sandhas equations \cite{AGS} in order to obtain electromagnetic
currents that are expressed in terms of two-body $t$-matrices $t$ and gauged
two-body $t$-matrices $t^\mu$. Our final results consist of gauge-invariant
expressions for the electromagnetic current of the three-particle bound state,
as well as the various electromagnetic transition currents of three particles.
Thus, in the case of three distinguishable nucleons, our expressions describe
the electromagnetic form factors of the bound \NNN\ system ($^3$H), and the
processes $Nd\rightarrow Nd\gamma$, $Nd\rightarrow \gamma N\!N\!N$, $\gamma
^3$H$\rightarrow Nd$, and $\gamma ^3$H$\rightarrow N\!N\!N$.
\begin{figure}[t]
\hspace*{.5cm} \epsfxsize=16cm\epsfbox{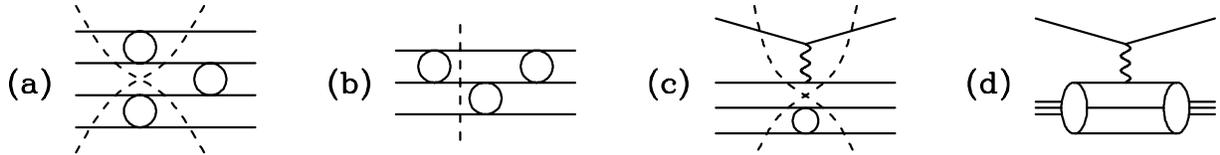}
\vspace{3mm}

\caption{\fign{overcount} Examples of last cuts (dashed lines) in relativistic
three- and four-body processes. (a) Ambiguous last cuts in the four-body
problem. (b) Unique last cut in the three-body problem. (c) Ambiguous last cuts
in electron scattering (top line) off three particles (bottom three straight
lines). (d) The traditional picture of elastic electron scattering off a
three-body bound state in impulse approximation.  Interpreted as a Feynman
graph (with the ellipses representing the bound state wave function), this
diagram is consistent with the graph of (c) being included twice, once for each
last cut shown. The diagram in (d) therefore contains overcounting.}
\end{figure}
\vspace{-5mm}

A notorious problem plaguing many four-dimensional approaches is the
overcounting of diagrams. For perturbation graphs, such overcounting can be
corrected simply by subtracting the overcounted terms. However, when
overcounting occurs within the scattering integral equations themselves, the way
to solve the overcounting problem becomes highly non-trivial. Indeed, only
recently has such an overcounting problem been solved in the context of
four-dimensional integral equations for the \NNpiNN\ system \cite{KB4d,PA4d}.

Overcounting can arise when reducible diagrams have ambiguous last cuts
\cite{KB4d,PA,Taylor}. An example of an ambiguous last cut for the four-body
problem is given in \fig{overcount}(a). Luckily, there is no overcounting in the
four-dimensional scattering equations considered here. This is because of the
purely three-body nature of our system (there is no coupling to two-body
channels) which makes all last cuts unique - see \fig{overcount}(b).  However,
as soon as coupling to an external photon is made, the system effectively
obtains coupling to the four-body sector and overcounting again becomes a
possibility. To see this explicitly, consider the calculation of electron
scattering off a three-body system in the relativistic impulse
approximation. One contribution to this process is given in
\fig{overcount}(c). The diagram shown contains ambiguous last cuts of the same
type as in \fig{overcount}(a). This ambiguity needs to be carefully taken into
account when summing the full perturbation series for the electron scattering
process. For example, to find the expression describing electron scattering off
a bound three-body system in the relativistic impulse approximation, one cannot
simply use the direct generalisation of the non-relativistic expression as given
diagrammatically in \fig{overcount}(d).  Interpreted as a Feynman graph, this
diagram overcounts interactions between the two spectator particles (the lowest
two lines in the diagram) since such interactions are contained in both the
initial and final three-body bound-state wave functions. This overcounting
corresponds to the diagram of \fig{overcount}(c) being included twice, once for
each last cut shown. Just this type of overcounting appears to be present in
Rupp and Tjon's calculation of the electromagnetic form factors of \triton.

An important feature of the gauging of equations method is that it not only
attaches photons everywhere in the three-body amplitude, but it also does this
without introducing any overcounting. Indeed, it is found that the gauging
procedure itself gives rise to subtraction terms that effectively remove all
overcounted contributions. In this way the complications brought about by
ambiguous last cuts like that of \fig{overcount}(c) are taken care of
automatically.

Preliminary results of the present paper (I) and the following paper (II) were
first reported a few years ago \cite{previous}. In the meantime we have applied
the gauging of equations method to the spectator formalism of Gross \cite{Gross}
to generate gauge invariant three-dimensional expressions describing any
hadronic or quark system interacting with an external electromagnetic probe
\cite{G3d}. In particular, we have derived the three-dimensional expressions for
the various electromagnetic transitions currents of the three-nucleon system
within the spectator approach \cite{nnn3d}. In this sense, the results of
Ref.~\cite{nnn3d} can be considered as a gauge-invariant three-dimensional
reduction of the four-dimensional results presented in I and II.  We have
also applied the gauging of equations method to the \NNpiNN\ system \cite{gpinn}
where, as previously mentioned, the overcounting of diagrams provides an extra
degree of complexity. For these works the present paper together with II form
the basic theoretical foundation, and provide the references where all
the missing details are given.

Although the power of the gauging of equations method is well demonstrated on
the example of the three-body system considered here, it should be emphasized
that the same method is just as easily applied to other strongly interacting
systems (quark or hadron) including those where the number of particles is not
conserved. Indeed this method has recently been used to gauge the $\pi N$ system
\cite{Haber}, and as mentioned above, the \NNpiNN\ system \cite{gpinn}.

As the gauging of equations method couples one external photon everywhere in the
strong interaction model, it also forms the basis for exact descriptions of more
complicated electromagnetic processes.  For example, we have recently shown how
to use this method to incorporate an {\em internal} photon into all possible
places in a strongly interacting system \cite{gg4d}. The resulting expressions
provide a way to calculate the complete set of lowest order electromagnetic
corrections to any strong interaction model described by integral equations.
Being complete, these electromagnetic corrections are therefore gauge invariant.
The gauging method can also be used to describe processes with more than one
photon.  For example, by gauging a strong interaction scattering equation twice,
we would obtain gauge invariant expressions for the corresponding Compton
scattering process.

Finally, it is important to note that although we are concerned in this paper
with the electromagnetic interaction for which gauge invariance (or current
conservation) is a major issue, the gauging of equations method itself is
totally independent of the type of external field involved. Thus all the
expressions for transition currents developed in this paper hold also for cases
where the external field is due to strongly or weakly interacting probes for
which current is not conserved (of course the exact form of the gauged inputs
would need to be chosen appropriately).





\section{GAUGING TWO DISTINGUISHABLE PARTICLES}

\subsection{Gauging the two-particle Green function \mbf{$G$}}

Gross and Riska \cite{GR} have shown how to construct a conserved current for
the two-particle system described by the Bethe-Salpeter (BS) equation. In
addition to the one-body current, which in general is not conserved, one also
needs the two-body interaction current obtained by attaching a photon to each
place inside all the Feynman diagrams defining the BS kernel. Such a current
satisfies the two-body Ward-Takahashi (WT) identity \cite{Bentz} corresponding
to the given model of strong interactions. On mass shell, this identity
guarantees that the matrix element of the current operator satisfies the
continuity equation. To obtain their result, Gross and Riska applied the
one-body WT identities \cite{WT} to a number of meson exchange diagrams of the
BS kernel. Here we would like to introduce a different method, that of gauging
equations, by rederiving the Gross-Riska result for the case of two particles.
In the subsequent sections, we shall illustrate the general nature of our method
by applying it to the case of the three-particle system.

The two-body WT identity \cite{Bentz} can be written as
\bea
\lefteqn{
q_\mu G^\mu (k_1k_2;p_1p_2)=
i[e_1G(k_1-q,k_2;p_1p_2)+e_2G(k_1,k_2-q;p_1p_2)}\hspace{3.3cm}\nn
&&
-G(k_1k_2;p_1+q,p_2)e_1-G(k_1k_2;p_1,p_2+q)e_2]  \eqn{WT2b}
\eea
where $p_1p_2$ ($k_1k_2$) are initial (final) momenta of the particles, the
photon is taken to be incoming with momentum $q$ so that $k_1+k_2=p_1+p_2+q$,
$G$ is the full two-body Green function given by
\bea
\lefteqn{ (2\pi)^4 \delta^4(p'_1+p'_2-p_1-p_2)G(p'_1p'_2;p_1p_2)=\int
d^4y_1d^4y_2d^4x_1d^4x_2
}\hspace{3cm}\nn
&&
e^{i(p'_1y_1+p'_2y_2-p_1x_1-p_2x_2)}\,\la 0|T\Psi^{(1)}(y_1) \Psi^{(2)}(y_2)
\bar{\Psi}^{(1)}(x_1)\bar{\Psi}^{(2)}(x_2)|0\ra      ,   \eqn{G2b}
\eea
and $G^\mu$ is the corresponding five-point function
\bea
\lefteqn{
G^\mu (k_1k_2;p_1p_2)= \int d^4y_1d^4y_2d^4x_1d^4x_2  } \hspace{2cm} \nn
&&
e^{i(k_1y_1+k_2y_2-p_1x_1-p_2x_2)}\,\la 0|T
\Psi^{(1)}(y_1)\Psi^{(2)}(y_2)\bar{\Psi}^{(1)}(x_1)
\bar{\Psi}^{(2)}(x_2)J^\mu(0)|0\ra     .    \eqn{G^mu2b}
\eea
In the above, $\Psi^{(i)}$ and $\bar{\Psi}^{(i)}$ are Heisenberg fields of
particle $i$, $T$ is the time ordering operator, $ J^\mu$ is the quantised
electromagnetic current operator, $|0\ra$ the physical vacuum, and $e_i$ is the
charge of the $i$-th particle. If the particles are isotopic doublets then $e_i$
includes an isospin factor, e.g.\ for nucleons
$e_i=\frac{1}{2}[1+\tau_3^{(i)}]e_p$ where $\tau_3$ is the Pauli matrix for the
third component of isospin, and $e_p$ is the charge of the proton.

For local field theory with distinguishable particles the electromagnetic
current operator is given by
\be
J^\mu(x) = -i \sum_{i=1}^{2}\frac{\partial{\cal L}}
{\partial\left(\partial_\mu\Psi^{(i)}\right)}e_i \Psi^{(i)}
+ J^\mu_{\mbox{\scriptsize other}}(x)
\ee
where $J^\mu_{\mbox{\scriptsize other}}(x)$ consists of partial derivatives with
respect to all other fields that are present in the Lagrangian ${\cal L}$. Note
that we have used translational invariance to write \eq{G^mu2b} in terms of
$J^\mu(0)$, and we have defined both $G$ and $G^\mu$ to be without total
4-momentum conserving $\delta$-functions. In \eq{WT2b} we have used a notation
where the two-body Green function $G$ is labelled by four momentum variables
even though only three are independent.  Such notation allows us to write down
the WT identity in an especially simple form that easily generalises to any
numbers of particles.
\begin{figure}[t]
\hspace*{3cm} \epsfxsize=10.7cm\epsfbox{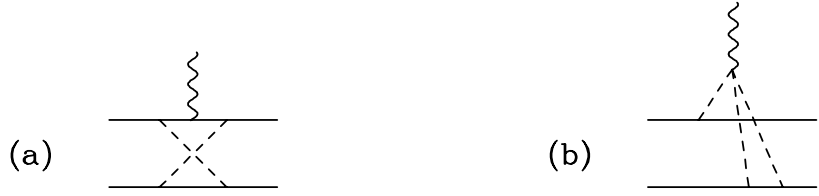}
\vspace{5mm}
\caption{Examples of the two types of Feynman diagram making up $G^\mu$.  (a) A
diagram that can be constructed by attaching a photon to a strong interaction
diagram of $G$. (b) A diagram that cannot be constructed by attaching a photon
to a diagram of $G$.}  \fign{g_mufig}
\end{figure}

One can classify the contributions to $G^\mu$ into two groups \cite{Woloshyn}.
The first of these consists of all Feynman diagrams that can be constructed from
$G$ by attaching a photon to an appropriate diagram of $G$. An example is given
in \fig{g_mufig}(a).  The second group consists of diagrams that cannot be
obtained from $G$ by attaching a photon, an example of which is given in
\fig{g_mufig}(b). The special feature of the second group is that each
contributing Feynman diagrams satisfies gauge invariance all on its own.

The goal of this paper is to show how to construct a gauge invariant $G^\mu$ by
attaching a photon to all possible places in every Feynman diagram contributing
to $G$. We shall refer to this attaching of photons everywhere as the {\em
gauging of} $G$. On the other hand, diagrams like that of \fig{g_mufig}(b) are
current conserving from the outset, and their construction is a separate problem
which will not be considered here. Thus for the purposes of this paper, we take
$G^\mu$ to be the result of the gauging of $G$, with the understanding that the
neglected contributions can always be added separately without affecting the
question of gauge invariance.

In a similar way, we use a superscript $\mu$ to indicate the vector quantity
obtained by attaching photons everywhere to an amplitude or potential. However
in this case we will require that {\em no photons be attached to external lines}
of Feynman diagrams making up the amplitude; nevertheless, we shall still refer
to the process of attaching the photons to all other places in the amplitude as
{\em gauging}.

The WT identity is usually given in terms of the exact strong interaction Green
functions of the underlying field theory. It follows that this identity must
also be valid to any order with respect the strong interaction coupling
constant. Here we would like to point out that the WT identity is also valid for
any particular single diagram of the strong interaction. This is because one can
always construct a Lagrangian with respect to which the diagram in question
represents the only case of some given order of the strong interaction. In this
case, the $G$ in the WT identity is given by just one diagram of the strong
interaction, and $G^\mu$ is the sum of the diagrams obtained from $G$ by
attaching the photon everywhere.  In the same way, the WT identity is valid for
any sum of diagrams included in $G$. Having this in mind, we consider $G$ to be
given by the Bethe-Salpeter equation where the kernel $V$ is given by a model
consisting of any number of connected two-particle irreducible Feynman diagrams.

We start by expressing $G$ in terms of its fully disconnected part $G_0$, and
the kernel $V$:
\be G=G_0+G_0VG . \eqn{G}
\ee
This is a symbolic equation that, for the case of two-particle scattering,
represents a shorthand notation for
\be
G(p'_1p'_2;p_1p_2) = G_0(p'_1p'_2;p_1p_2) +
\int \frac{d^4r_1}{(2\pi)^4}\frac{d^4s_1}{(2\pi)^4} \,
G_0(p'_1p'_2;r_1r_2) V(r_1r_2;s_1s_2) G(s_1s_2;p_1p_2) \eqn{Glong}
\ee
where it is understood that $p'_1+p'_2=p_1+p_2=r_1+r_2=s_1+s_2$.  In \eq{Glong},
neither the Green functions nor the kernel contain $\delta$-functions
corresponding to total momentum conservation. Thus the disconnected Green
function $G_0$ contains only one $\delta$-function and can be written as
\be
G_0(p'_1p'_2;p_1p_2) = (2\pi)^4 \delta^4(p'_1-p_1) d_1(p_1) d_2(p_2)
= (2\pi)^4 \delta^4(p'_2-p_2) d_1(p_1) d_2(p_2) \eqn{G0}
\ee
where $d_i$ is the dressed propagator of particle $i$. To save on notation we
write \eq{G0} symbolically as
\be
G_0 = d_1 d_2    \eqn{G0_short}
\ee
where momentum labels and the momentum conserving $\delta$-function [together
with factor $(2\pi)^4$] have been suppressed.

\eq{G} is basically a topological statement regarding the two-particle
irreducible structure of Feynman diagrams belonging to $G$; as such, it can be
utilised directly to express the structure of the same Feynman diagrams, but
with photons attached everywhere. Thus from \eq{G} it immediately follows that
\be
G^\mu =G_0^\mu + G_0^\mu VG+G_0V^\mu G+G_0VG^\mu .  \eqn{G^mu}
\ee
This result expresses $G^\mu$ in terms of an integral equation, and illustrates
what we mean by {\em gauging an equation}, in this case the gauging of \eq{G}.
Implied in \eq{G^mu} is the result
\be
[G_0VG]^\mu = G_0^\mu VG+G_0V^\mu G+G_0VG^\mu ,  \eqn{product}
\ee
which illustrates a rule for the gauging of products that is rather reminiscent
of the product rule for derivatives. Although \eq{product} follows from a
topological argument, we shall postulate such a product rule to hold also in
other cases where topological arguments may not be applicable. Again, \eq{G^mu}
is a symbolic equation representing
\bea
\lefteqn{
G^\mu(k_1k_2;p_1p_2) = G^\mu_0(k_1k_2;p_1p_2) + 
\int \!\frac{d^4r_1}{(2\pi)^4}\frac{d^4s_1}{(2\pi)^4} \,
G^\mu_0(k_1k_2;r_1r_2) V(r_1r_2;s_1s_2) G(s_1s_2;p_1p_2)} \hspace{4cm} \nn
&&+ \int \!\frac{d^4t_1}{(2\pi)^4}\frac{d^4u_1}{(2\pi)^4}\,
G_0(k_1k_2;t_1t_2) V^\mu(t_1t_2;u_1u_2) G(u_1u_2;p_1p_2) \nn
&&+ \int \!\frac{d^4v_1}{(2\pi)^4}\frac{d^4w_1}{(2\pi)^4}\,
G_0(k_1k_2;v_1v_2) V(v_1v_2;w_1w_2) G^\mu(w_1w_2;p_1p_2) 
\hspace{1cm} \eqn{G^mulong}
\eea
where now the presence or absence of a photon with momentum $q$ needs to be
taken into account in specifying the momentum conservation relations:
$k_1+k_2=p_1+p_2+q$, where $k_1+k_2=t_1+t_2=v_1+v_2=w_1+w_2$, and
$p_1+p_2=r_1+r_2=s_1+s_2=u_1+u_2$.

In \eq{G^mu}, both $G^\mu$ and $G_0^\mu$ are obtained from the Green functions
$G$ and $G_0$, respectively, by attaching photons everywhere. It is
therefore important to note that the gauged potential $V^\mu$ is similarly
obtained from $V$, but with no photons attached to external legs. This is
because such contributions are already taken into account in the terms
$G_0^\mu VG$ and $G_0VG^\mu$ of \eq{G^mu}.
\eqs{G} and (\ref{G^mu}) are a set of linear integral equations for $G$
and $G^\mu$, and could be solved as such if $V$ and $V^\mu$ were given. However,
we can also formally solve the equation for $G^\mu$, and thus express it
directly in terms of $G$. Simple algebra gives:
\ben
G^\mu -G_0VG^\mu =G_0^\mu +G_0^\mu VG+G_0V^\mu G,
\een
\be
G^\mu =(I-G_0V)^{-1}\left[ G_0^\mu (1+ VG)+G_0V^\mu G\right]
=G\left(G_0^{-1}G_0^\mu G_0^{-1}+ V^\mu \right)G         \eqn{G^mutemp}
\ee
where \eq{G} was used in the last step. Defining the electromagnetic vertex
function $\Gamma^\mu$ by
\be
G^\mu =G\Gamma ^\mu G,         \eqn{Gamma^mudef}
\ee
\eq{G^mutemp} gives the essential result of this section:
\be
\Gamma^\mu =\Gamma_0^\mu+V^\mu  \eqn{Gamma^mu}
\ee
where
\be
\Gamma_0^\mu = G_0^{-1}G_0^\mu G_0^{-1}.    \eqn{Gamma_0^mu_def}
\ee
By gauging \eq{G0_short} one obtains
\be
G_0^\mu = d_1^\mu d_2 +  d_1 d_2^\mu
\ee
so that
\be
\Gamma_0^\mu = 
\ggama_1^\mu d_2^{-1} + d_1^{-1} \ggama_2^\mu \eqn{Gamma_0^mu}
\ee
where
$\ggama_i^\mu$ is the one-particle electromagnetic vertex function defined
by the equation
\be
d_i^\mu = d_i\, \ggama_i^\mu d_i.
\ee
Note that for nucleons, to lowest order in the strong interaction, $\ggama_i^\mu
= e_i\gamma^\mu$ where $\gamma^\mu$ is a Dirac matrix. In \eq{Gamma^mu}
$\Gamma_0^\mu$ is thus the sum of single particle currents, and $V^\mu$ is the
two-body interaction current of the given model. \eq{Gamma^mu} is illustrated in
\fig{gamma_2d}.  This is the essential result of Gross and Riska, and has been
derived here using the gauging of equations method.
\begin{figure}[t]
\hspace*{3cm}  \epsfxsize=10cm\epsfbox{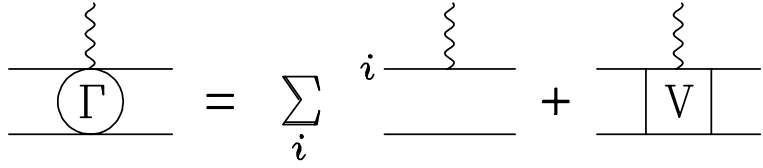}
\vspace{2mm}
\caption{Illustration of \protect\eq{Gamma^mu} expressing the two-particle
electromagnetic vertex function $\Gamma^\mu$ as a sum one-body currents and 
the two-particle interaction current.}
\fign{gamma_2d}
\end{figure}

\subsection{Ward-Takahashi identity for \mbox{\boldmath $G^\mu$}}

The validity of \eq{G^mu} for the gauging of $G$ is clear from the topological
argument given above. We would nevertheless like to show explicitly that the
$G^\mu$, constructed in this way, satisfies the WT identity. To do this, we
first prove that the gauged potential $V^\mu$ satisfies the WT identity even
though no photons are attached to its external legs. Using a shorthand notation
defined by
\be
G(k_1k_2;p_i+q) = \left\{\begin{array}{ll} G(k_1k_2;p_1+q,p_2) & (i=1) \\
G(k_1k_2;p_1,p_2+q) & (i=2)\end{array} \right .       \eqn{short1}
\ee
\be
G(k_i-q;p_1p_2) = \left\{\begin{array}{ll} G(k_1-q,k_2;p_1p_2) & (i=1) \\
G(k_1,k_2-q;p_1p_2) & (i=2)\end{array} \right .       \eqn{short2}
\ee 
we can write the WT identities for $G_0^\mu$ and the quantity
$[G_0VG_0]^\mu$ as
\be
q_\mu G_0^\mu(k_1k_2;p_1p_2)=i\sum_{i=1}^2[e_iG_0(k_i-q;p_1p_2)-
G_0(k_1k_2;p_i+q)e_i],    \eqn{qG0}
\ee
\be
q_\mu [G_0VG_0]^\mu(k_1k_2;p_1p_2) =i\sum_{i=1}^2[e_iG_0VG_0(k_i-q;p_1p_2)
-G_0VG_0(k_1k_2;p_i+q)e_i].     \eqn{qG0VG0}
\ee
Note that $[G_0VG_0]^\mu$ satisfies the WT identity because, unlike $V^\mu$, it
has photons attached everywhere.  Although \eq{qG0} and \eq{qG0VG0} would be
automatically true if field theory were being solved exactly, here we work
within a strong interaction model specified by the input quantities $V$ and
$G_0$; as such, these equations are assumed to be true by construction.

Using the product rule of \eq{product}, we have that
\bea
\lefteqn{q_\mu [G_0VG_0]^\mu(k_1k_2;p_1p_2) = 
q_\mu G_0^\mu VG_0(k_1k_2;p_1p_2) +G_0\,q_\mu V^\mu G_0(k_1k_2;p_1p_2)
}\hspace{8cm}\nn
&&+G_0Vq_\mu G_0^\mu(k_1k_2;p_1p_2) .\hspace{.5cm}
\eea
Writing out the integrals implied by this expression [see \eqs{Glong} and
(\ref{G^mulong})] and using \eqs{qG0}, (\ref{qG0VG0}) and (\ref{G0}), we obtain
that
\be
q_\mu V^\mu(k_1k_2;p_1p_2)
=i\sum_{i=1}^2\left[e_iV(k_i-q;p_1p_2)-V(k_1k_2;p_i+q) e_i\right] \eqn{qVmu}
\ee
which is the WT identity for $V^\mu$. We may now use \eqs{qVmu} and
(\ref{qG0}) together with \eq{Gamma^mu} to evaluate
$q_\mu \Gamma^\mu(k_1k_2;p_1p_2)$. Taking into account that
$G_0^{-1}-V=G^{-1}$, we obtain
\be
q_\mu \Gamma^\mu(k_1k_2;p_1p_2)=i\sum_{i=1}^2\left[G^{-1}(k_1k_2;p_i+q)e_i
-e_iG^{-1}(k_i-q;p_1p_2)\right]  .  \eqn{Gamma-wti}
\ee
Using this result and \eq{Gamma^mudef} it immediately follows that
\be
q_\mu G^\mu(k_1k_2;p_1p_2)=i\sum_{i=1}^2\left[e_iG(k_i-q;p_1p_2)-
G(k_1k_2;p_i+q)e_i\right],    \eqn{qG}
\ee
thus proving the WT identity for the $G^\mu$ obtained by the
gauging of equations method.

\subsection{Gauging the two-body bound-state wave function \mbf{$\psi$}}

So far we have defined ``gauging'' to be the process where photons are attached
to all places in perturbation diagrams. As Green functions and potentials have a
diagrammatic interpretation, the gauging of these quantities has therefore a
clear meaning.  On the other hand, the bound-state wave function is a purely
nonperturbative quantity, and thus cannot be associated with perturbation
diagrams. One can nevertheless define the gauged bound-state wave function
$\psi^\mu$ by formally gauging the bound-state Bethe-Salpeter equation using our
product rule. Thus by gauging the equation
\be
\psi = G_{0}V\psi \eqn{psi}
\ee
we obtain
\be
\psi^\mu = \left(G_{0}V\right)^\mu\psi+G_{0}V\psi^\mu   \eqn{psi^mu}
\ee
in this way defining $\psi^\mu$.
Here the symbol $\psi$ represents the two-body bound-state wave function
$\psi_P(p_1p_2)$ defined by
\be
(2\pi)^4 \delta^4(P-p_1-p_2)\psi_P(p_1p_2) = \int d^4x_1d^4x_2
e^{i(p_1x_1+p_2x_2)} \la 0|T\Psi^{(1)}(x_1)\Psi^{(2)}(x_2)|P\ra  \eqn{psi_2b}
\ee
where $P^2=m^2$ with $m$ being the mass of the bound two-body system. To save on
notation, we do not write explicit spin and isospin labels in the state $|P\ra$
nor in the wave function $\psi_P$; nevertheless, such labels (and if necessary,
sums over such labels) are to be understood as implicitly present.

It is now straightforward to show that the $\psi^\mu$ defined by \eq{psi^mu} is
just the quantity $\psi^\mu(k_1k_2;P)$ given by
\be
\psi^\mu(k_1k_2;P) = \int d^4x_1d^4x_2
e^{i(k_1x_1+k_2x_2)} \la 0|T\Psi^{(1)}(x_1)\Psi^{(2)}(x_2)J^\mu(0)|P\ra 
\eqn{psi^mu_2b}
\ee
where $k_1+k_2=P+q$. The proof of this proceeds as follows. Assuming the field
theory under consideration admits a two-body bound state, the Green function
$G$, defined by \eq{G2b}, exhibits a pole at $P^2=m^2$ where $P$ is the total
four-momentum of the bound two-body system. Indeed from \eq{G2b} it can be shown
that as $P^2\rightarrow m^2$
\be
G(p_1'p_2',p_1p_2) \sim i\, 
\frac{\psi_P(p_1'p_2')\bar{\psi}_P(p_1p_2)}{P^2-m^2}
\ee
where $P=p_1+p_2=p'_1+p'_2$ and $\psi_P$ is given by \eq{psi_2b}.
Similarly from \eq{G^mu2b} one can show that for $P^2\rightarrow m^2$, 
\be
G^\mu(k_1k_2,p_1p_2) \sim i\, 
\frac{\psi^\mu(k_1k_2;P)\bar{\psi}_P(p_1p_2)}{P^2-m^2}. \eqn{G^mu-pole}
\ee
with $\psi^\mu(k_1k_2;P)$ given by \eq{psi^mu_2b}.  Using the last two results
in \eq{G^mu}, equating the residues at $P^2=m^2$, and writing $\psi^\mu$ as
shorthand for $\psi^\mu(k_1k_2;P)$, we obtain that
\be
\psi^\mu\bar{\psi} = \left(G_0V\right)^\mu\psi\bar{\psi}
+ G_0V\psi^\mu\bar{\psi}.
\ee
We therefore deduce that the $\psi^\mu$ given by \eq{psi^mu_2b} does indeed
satisfy \eq{psi^mu}.
\eqs{psi} and (\ref{psi^mu}) form a coupled set of equations which can
be formally solved for $\psi^\mu$ giving
\bea
\psi^\mu \!&=&\! \left(1-G_{0}V\right)^{-1}\left(G_{0}V\right)^\mu\psi
= GG^{-1}_{0}\left(G_{0}V\right)^\mu\psi\nn
\!&=&\! G \left( G^{-1}_{0}G^\mu_{0}G^{-1}_{0}+V^\mu\right) \psi .
\eea
Recalling \eq{Gamma^mu} we obtain
\be
\psi^\mu = G \Gamma^\mu \psi.   \eqn{psi^mu_2}
\ee
Comparison with \eq{Gamma^mudef} shows that $\psi^\mu$ can be obtained from
$G^\mu$ by taking the right-hand residue at the two-body bound-state pole [this
of course is also obvious from \eq{G^mu-pole}].  As such, $G_0^{-1}\psi^\mu$ is
just the transition current describing the photo/electrodisintegration of the
two-body bound state, as discussed shortly.

It is often convenient to work with the bound-state vertex function $\phi$
defined by the equations
\be
\psi = G_0\phi\hspace{5mm};\hspace{5mm} \bar{\psi} = \bar{\phi}G_0. \eqn{phi}
\ee
To gauge $\phi$ we follow the same idea as for the bound-state wave function,
namely, we define $\phi^\mu$ by formally gauging the bound-state equation for
$\phi$ [which follows from \eq{psi}]. In this way we obtain the coupled set of
equations
\bea
\phi \!&=&\! VG_0\phi \eqn{SE_phi}  \\
\phi^\mu \!&=&\! \left(VG_0\right)^\mu\phi + VG_0\phi^\mu
\eea
which can be solved as above to give
\be
\phi^\mu = G_0^{-1}G\left(VG_0\right)^\mu\phi .  \eqn{phi^mu}
\ee
Writing $(VG_0)^\mu=V^\mu G_0+VG_0^\mu$ and $G_0^{-1}G=1+TG_0$ where $T$ is the
$t$-matrix defined by
\be
G=G_0+G_0TG_0 ,       \eqn{t_def}
\ee
\eq{phi^mu} reduces after some simple algebra to the result
\be
\phi^\mu = (1+TG_0)\Gamma^\mu\psi - \Gamma_0^\mu\psi. \eqn{phi^mu_final}
\ee

\subsection{Ward-Takahashi identities for \mbf{$\psi^\mu$} and \mbf{$\phi^\mu$}}

The WT identity for $\psi^\mu$ follows from \eq{psi^mu_2} and the WT identity
for $\Gamma^\mu$, \eq{Gamma-wti}. Writing \eq{psi^mu_2} in its full numerical
form
\be
\psi^\mu(k_1k_2;P) = \int \frac{d^4t_1}{(2\pi)^4}\frac{d^4p_1}{(2\pi)^4}
\,G(k_1k_2;t_1t_2)\Gamma^\mu(t_1t_2;p_1p_2) \psi(p_1p_2;P),
\ee
we may use \eq{Gamma-wti} to obtain
\bea
\lefteqn{q_\mu\psi^\mu(k_1k_2;P) = 
\int \frac{d^4t_1}{(2\pi)^4}\frac{d^4p_1}{(2\pi)^4}G(k_1k_2;t_1t_2)}
\hspace{2cm}\nn
&&\,i\sum_{i=1}^2\left[G^{-1}(t_1t_2;p_i+q)e_i-e_i G^{-1}(t_i-q;p_1p_2)\right]
\psi(p_1p_2;P).
\eea
Since $G^{-1}\psi_P=0$, the second term in the square brackets does not
contribute and the above equation reduces down to the WT identity for
$\psi^\mu$,
\be
q_\mu\psi^\mu(k_1k_2;P) =
i\left[e_1\psi(k_1-q,k_2;P)+e_2\psi(k_1,k_2-q;P)\right].
\eqn{psi-wti}
\ee

The WT identity for $\phi^\mu$ is found in a similar manner. Written out
numerically, \eq{phi^mu} is
\be
\phi^\mu(k_1k_2;P) = \int \frac{d^4t_1}{(2\pi)^4}\frac{d^4p_1}{(2\pi)^4}
\,[G_0^{-1}G](k_1k_2;t_1t_2) [VG_0]^\mu(t_1t_2;p_1p_2) \phi(p_1p_2;P).
\ee
Note that the square brackets here indicate the resultant function to which the
following momentum variables refer. As $[VG_0]^\mu$ is a gauged input, it
satisfies the Ward-Takahashi identity from the outset. We can therefore write
\bea
\lefteqn{q_\mu\phi^\mu(k_1k_2;P)=
\int \frac{d^4t_1}{(2\pi)^4}\frac{d^4p_1}{(2\pi)^4}\,
[G_0^{-1}G](k_1k_2;t_1t_2)}\hspace{2cm} \nn
&&i\sum_{i=1}^2\left\{ e_i[VG_0](t_i-q;p_1p_2)- [VG_0](t_1t_2;p_i+q)e_i
\right\} \phi(p_1p_2;P) .
\eea
The first term in the curly brackets can be integrated over $p_1$ using the fact
that $VG_0\phi=\phi$. Similarly the second term in the curly brackets can be
integrated over $t_1$ using the fact that $G_0^{-1}GVG_0=G_0^{-1}G-1$. Only one
term survives,
\be
q_\mu\phi^\mu(k_1k_2;P)=\int \frac{d^4p_1}{(2\pi)^4}\,i\sum_{i=1}^2 (2\pi)^4 
\delta^4(k_i-p_i-q) e_i \phi(p_1p_2;P) ,
\ee
which gives the Ward-Takahashi identity for $\phi^\mu$:
\be
q_\mu\phi^\mu(k_1k_2;P)= i \left[ e_1\phi(k_1-q,k_2;P)
+ e_2\phi(k_1,k_2-q;P)\right] .     \eqn{phi-wti}
\ee
Note that neither of the WT identities of \eq{psi-wti} nor \eq{phi-wti} are
zero, even for on mass shell $k_1$ and $k_2$. Thus neither $\psi^\mu$ nor
$\phi^\mu$ obey current conservation.

\subsection{Two-body electromagnetic transition currents}

To obtain the physical $t$-matrix for any reaction involving an electromagnetic
probe interacting with a quark or hadronic system, it is sufficient to specify
the corresponding matrix element of the electromagnetic current operator. We
refer to such a matrix element simply as an {\em electromagnetic transition
  current} and denote it using the symbol $j^\mu$.  Thus, for example, a
process like two-nucleon Bremsstrahlung $N\!N \rightarrow \gamma N\!N$ is
described by the $N\!N\rightarrow N\!N$ electromagnetic transition current
$j_{00}^\mu$ given by
\be
j_{00}^\mu = G_0^{-1}G^\mu G_0^{-1}.      \eqn{j_00}
\ee
The physical $t$-matrix is then found by contracting $j_{00}^\mu$ with the
photon polarisation vector $\varepsilon_\mu$.

For processes with one two-body bound state, like deuteron photodisintegration,
one determines the electromagnetic transition current $j_0^\mu$ from $G^\mu$ by
taking the appropriate residue at the bound-state pole.  Thus, to determine the
$(ij)\rightarrow ij$ electromagnetic transition current describing the process
$\gamma (ij)\rightarrow ij$ where a photon is absorbed on the bound state $(ij)$
producing free particles $i$ and $j$ in the final state, we take the residue of
\eq{Gamma^mudef} on the right to obtain
\bea j_0^\mu
&=& G_0^{-1} G \Gamma^\mu \psi \\ &=&\! (1+TG_0)\Gamma^\mu\psi
\eea
where $T$ is the off-shell two-body $t$-matrix. As noted previously, a
comparison with \eq{psi^mu_2} shows that the $(ij)\rightarrow ij$
electromagnetic transition current and the gauged bound-state wave function are
simply related by
\be
j_0^\mu = G_0^{-1} \psi^\mu.
\ee

It is also interesting to examine the relationship between $j_0^\mu$ and the
gauged bound-state vertex function $\phi^\mu$. From \eq{phi^mu_final}
one immediately obtains that
\be
j_0^\mu= \phi^\mu + \Gamma_0^\mu G_0\phi.
\ee
From this equation it is evident that the difference between $\phi^\mu$ and the
electromagnetic transition current $j_0^\mu$ is that $\phi^\mu$ has no photons
attached to external legs.  This suggests that $\phi^\mu$ does not conserve
current, while $j_0^\mu$ does. Indeed $q_\mu j_0^\mu = G_0^{-1} q_\mu \psi^\mu =
0$ for on mass shell external legs since the momentum shifts ($-q$) contained in
the WT identity for $\psi^\mu$, \eq{psi-wti}, ensure that the poles of
$q_\mu\psi^\mu$ do not cancel the zero of $G_0^{-1}$.

Lastly we consider the case where an electromagnetic probe interacts with two
particles which are bound in both initial and final states, for example elastic
electron-deuteron scattering. Such processes are described by the
electromagnetic bound-state current $j^\mu$ which can be found from
\eq{Gamma^mudef} by taking residues at both the initial and final bound-state
poles. As these bound states can have different total momenta, we write the
bound-state current as
\be
j^\mu = \bpsi_K \Gamma^\mu \psi_P
\ee
where $K^2=P^2=m^2$. Current conservation for $j^\mu$ is obtained
by using the WT identity for $\Gamma^\mu$, \eq{Gamma-wti}, and noting that
$\bpsi G^{-1}=G^{-1}\psi=0$.

\newpage
\section{GAUGING THREE DISTINGUISHABLE PARTICLES}

Having gauged the two-body system, we are ready to apply the gauging of
equations method to the substantially more complicated case of three strongly
interacting particles. 

\subsection{Gauging the three-particle Green function}

The Ward-Takahashi identity for three particles can be derived by following the
same procedure as used by Bentz \cite{Bentz} for the two-particle case. One
obtains
\bea 
\lefteqn{ q_\mu G^\mu (k_1k_2k_3;p_1p_2p_3)=
i[e_1G(k_1-q,k_2k_3;p_1p_2p_3)+e_2G(k_1,k_2-q,k_3;p_1p_2p_3)}
\hspace{3cm}\nn
&&+e_3G(k_1k_2,k_3-q;p_1p_2p_3)-G(k_1k_2k_3;p_1+q,p_2p_3)e_1\nn
&&-G(k_1k_2k_3;p_1,p_2+q,p_3)e_2-G(k_1k_2k_3;p_1p_2,p_3+q)e_3] .
\hspace{1cm} \eqn{WGmu}
\eea
By extending the shorthand notation of \eqs{short1} and (\ref{short2}) to the
case of three particles, the three-particle Ward-Takahashi identity can then be 
written more concisely as
\be
q_\mu G^\mu (k_1k_2k_3;p_1p_2p_3)=i\sum_{i=1}^3
[e_iG(k_i-q;p_1p_2p_3)-G(k_1k_2k_3;p_i+q)e_i].
\ee
Here $G$ is the full Green function of the three-particle system 
\bea
\lefteqn{(2\pi)^4\delta^4(p'_1+p'_2+p'_3-p_1-p_2-p_3)G(p'_1p'_2p'_3;p_1p_2p_3)=
\int d^4y_1d^4y_2d^4y_3d^4x_1d^4x_2d^4x_3} \hspace{1cm}\nn
&&e^{i(p'_1y_1+p'_2y_2+p'_3y_3-p_1x_1-p_2x_2-p_3x_3)}\,
\la 0|T\Psi^{(1)}(y_1)\Psi^{(2)}(y_2)\Psi^{(3)}(y_3)
\bar{\Psi}^{(1)}(x_1)\bar{\Psi}^{(2)}(x_2)\bar{\Psi}^{(3)}(x_3)|0\ra\nn
&& \eqn{G6pt}
\eea
and $G^\mu$ is the corresponding seven-point function 
\bea
\lefteqn{G^\mu (k_1k_2k_3;p_1p_2p_3)=
\int d^4y_1d^4y_2d^4y_3d^4x_1d^4x_2d^4x_3}
\hspace{0cm}\nn
&&e^{i(k_1y_1+k_2y_2+k_3y_3-p_1x_1-p_2x_2-p_3x_3)}
\la 0|T\Psi^{(1)}(y_1) \Psi^{(2)}(y_2)\Psi^{(3)}(y_3)
\bar{\Psi}^{(1)}(x_1)\bar{\Psi}^{(2)}(x_2)\bar{\Psi}^{(3)}(x_3)J^\mu(0)|0\ra.\nn
&&
\eqn{G7pt}
\eea

In practise, the Green function for three distinguishable particles is specified
by the integral equation
\be
G=G_0+G_0VG           \eqn{G3}
\ee
where the potential $V$ is three-particle irreducible, and in
the absence of three-body forces, is written as a sum of three disconnected
potentials $V_i$:
\be
V=V_1 + V_2 + V_3  \eqn{Vsum}.
\ee
Here we use the usual spectator notation of three-body theory: defining $(ijk)$
to be a cyclic permutation of $(123)$, $V_i$ is the potential where particles
$j$ and $k$ are interacting while particle $i$ is a spectator. Explicitly we
have that
\be
V_i(p'_1p'_2p'_3,p_1p_2p_3) =v_i(p'_jp'_k,p_jp_k) d_i^{-1}(p_i)
(2\pi)^4 \delta^4(p'_i-p_i)
\eqn{V_i}
\ee
where $v_i$ is the two-body potential between particles $j$ and $k$. Neither
$V_i$ nor $v_i$ contain total momentum conserving $\delta$-functions, and the
presence of $\delta^4(p'_i-p_i)$ in \eq{V_i} is due purely to the
disconnectedness of the potential $V_i$.  In keeping with our shorthand
notation, we write \eq{V_i} as
\be
V_i = v_i d_i^{-1}.  \eqn{V_i_short}
\ee

Our goal is to gauge \eq{G3} and in this way obtain the electromagnetic
transition currents for the three-body system. As the gauging procedure is
symbolically identical to that already presented for the two-body system, we
conclude that \eqs{Gamma^mudef}-(\ref{Gamma_0^mu_def}) apply also to the
three-body system. Thus the seven-point function can be written as
\be
G^\mu =G \Gamma^\mu G                                \eqn{G^mu3}
\ee
where the electromagnetic vertex function for three particles is given by
\be
\Gamma^\mu = G_0^{-1}G_0^\mu G_0^{-1} + V^\mu .     \eqn{Gamma^mu3}
\ee
In the three-particle case, however, $V$ is given by \eqs{Vsum} and
(\ref{V_i_short}) so that
\be
V^\mu=V_1^\mu + V_2^\mu + V_3^\mu \eqn{Vmusum}.
\ee
To obtain $V_i^\mu$ all that is needed is to gauge the $V_i$ of \eq{V_i_short}.
Yet because of the presence of the inverse propagator $d_i^{-1}$ in $V_i$, there
is some question of how to do this consistently. An unambiguous answer is
provided by gauging \eq{G3} with $V$ explicitly given by \eq{Vsum}
and \eq{V_i_short}. In this case we write \eq{G3} as
\be
G=G_0 + \sum_i G_0 V_i  G = G_0 + \sum_i D_{0i} v_i  G  \eqn{G4}
\ee
where $D_{0i}$ is the fully disconnected two-particle propagator defined by
\be
D_{0i} = d_j d_k.
\ee
The gauging of \eq{G4} thus involves the term
\be
(G_0V_iG)^\mu = (D_{0i} v_i G)^\mu =
\left(D_{0i}^\mu v_i+D_{0i}v_i^\mu\right)G+D_{0i}v_iG^\mu.
\ee
Now using the fact that $D_{0i}^\mu = (G_0^\mu-D_{0i}d_i^\mu)d_i^{-1}$, the
above equation gives
\bea
(G_0V_iG)^\mu&=&\left[G_0^\mu V_i-G_0\ggama_i^\mu v_i+G_0\left(v_i^\mu d_i^{-1}
\right) \right]G+G_0V_iG^\mu\nn
&=&G_0^\mu V_iG + G_0\left(-\ggama_i^\mu v_i+v_i^\mu d_i^{-1}\right)G
+G_0V_iG^\mu
\eea
from which it follows that
\be
V_i^\mu = v_i^\mu d_i^{-1}-v_i \hggama_i^\mu. \eqn{Vmu}
\ee
It is seen that the same result would follow from gauging \eq{V_i_short}
directly using our product rule, as long as we use the prescription
\be
\left(d_i^{-1}\right)^\mu  = -\ggama_i^\mu.    \eqn{d-1}
\ee
This prescription can also be obtained by formally gauging the
identity $d_id_i^{-1}=1$ and taking $1^\mu=0$.  The expression for the gauged
potential $V_i^\mu$, \eq{Vmu}, is illustrated in \fig{Vimu}. The negative sign
in front of the term $v_i \hggama_i^\mu$ may appear to be surprising, yet it is
just what is needed to stop overcounting. Consider for example the gauging of
the term $G_0VG$ appearing in \eq{G3}: $(G_0VG)^\mu=G_0^\mu VG+G_0V^\mu
G+G_0VG^\mu$. It is apparent that the rightmost diagram of \fig{Vimu} appears in
each of the three terms $G_0^\mu VG$, $G_0V^\mu G$, and $G_0VG^\mu$. Thus the
negative sign in question is needed to ensure that this diagram contributes only
once to the gauging of $G_0VG$.
\begin{figure}[t]
\hspace*{3cm}  \epsfxsize=9cm\epsfbox{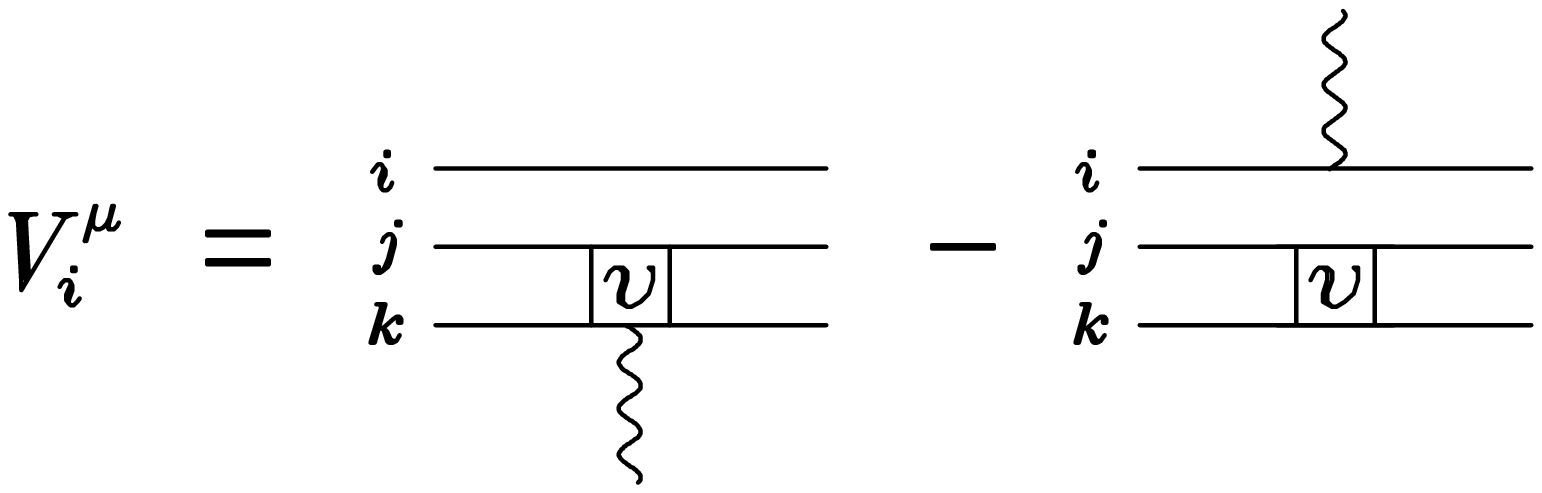}
\caption{\fign{Vimu} Illustration of \protect\eq{Vmu} for the gauged
disconnected potential $V_i^\mu$.}
\end{figure}

The fully disconnected three-particle Green function $G_0$ is given by
\be
G_0 = d_i d_j d_k \eqn{G_0ijk}
\ee
where two momentum conserving $\delta$-functions are implied in the used
shorthand notation. Gauging this expressions gives
\be
G_0^\mu =\sum_{i=1}^3 d_i^\mu d_j d_k
        =\sum_{i=1}^3 d_i^\mu D_{0i}            \eqn{G_0^mu}
\ee
\begin{figure}[b]
\hspace*{1cm}  \epsfxsize=14cm\epsfbox{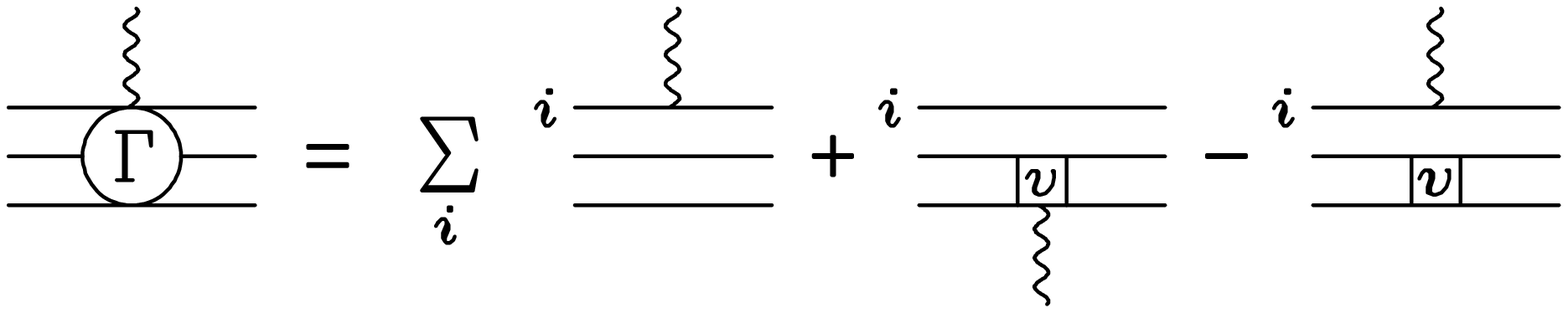}
\caption{\fign{gamma_3d} Illustration of \protect\eq{hello} expressing the
three-particle electromagnetic vertex function $\Gamma^\mu$ as a sum of one- and
two-particle currents.}
\end{figure}
where the sum is over the three cyclic permutations of ($ijk$). The impulse term
in \eq{Gamma^mu3} can then be written as
\be
G_0^{-1}G_0^\mu G_0^{-1} = \sum_{i=1}^3 \hggama_i^\mu D_{0i}^{-1} .
\eqn{impulse_term}
\ee
Using this and \eq{Vmu} in \eq{Gamma^mu3} then gives
\be
\Gamma^\mu=\sum_{i=1}^3 \left(\hggama_i^\mu D_{0i}^{-1}+v_i^\mu d_i^{-1}
-v_i\hggama_i^\mu \right)  \eqn{hello}
\ee
which we illustrate in \fig{gamma_3d}.

Noting that the full  two-body Green function $D_i$ satisfies the equation
\be
D_i = D_{0i} +
D_{0i} v_i D_i,
\ee
we can also write \eq{hello} as
\be
\Gamma^\mu=\sum_{i=1}^3 \left(\hggama_i^\mu D_i^{-1}+v_i^\mu d_i^{-1}
\right)  .     \eqn{vmu}
\ee
\eq{vmu} [or equivalently \eq{hello}] is the main result of this section and
extends the Gross-Riska result of \eq{Gamma^mu} to the three-particle
sector. It gives the precise way that the one- and two-body currents need to
combine in order to obtain proper gauge invariance.

\subsection{Three-body bound-state current}

The three-body bound-state current $j^\mu$ describes processes where a system of
three strongly interacting particles is bound both before and after the
interaction with an electromagnetic probe. Examples include elastic electron
scattering from $^3$H and $^3$He \cite{Amroun}, and elastic electron scattering
from the proton considered as a bound three-quark system \cite{Sill}.

The bound-state current $j^\mu$ is found by taking left and right residues of
\eq{G^mu3} at the initial and final three-particle bound-state poles.
With $M$ being the mass of the bound state, it can be shown that
\be
G(p'_1p'_2p'_3;p_1p_2p_3) \sim i \frac{\Psi_P(p'_1p'_2p'_3)
\bar{\Psi}_P(p_1p_2p_3)}{P^2-M^2} \hspace{1cm}\mbox{as} 
\hspace{5mm} P^2\rightarrow M^2                     \eqn{Gpole}
\ee
where $\Psi_P$ is the three-particle bound-state wave function defined by
\bea
\lefteqn{(2\pi)^4 \delta^4(P-p_1-p_2-p_3)\Psi_P(p_1p_2p_3)  = }\hspace{2cm} \nn
&& \int d^4x_1d^4x_2d^4x_3 e^{i(p_1x_1+p_2x_2+p_3x_3)}
\la 0|T\Psi^{(1)}(x_1)\Psi^{(2)}(x_2)\Psi^{(3)}(x_3)|P\ra .      \eqn{Psi_P}
\eea
Here $|P\ra$ is the eigenstate of the full Hamiltonian corresponding to the
three-particle bound state with momentum $P^\mu$. We thus find that
\bea
j^\mu &=&\bar{\Psi}_{K}\Gamma ^\mu \Psi_P \\ \eqn{j^mu1} &=&
\sum_{i=1}^3\bar{\Psi}_{K}\left(\hggama_i^\mu D_{0i}^{-1}+v_i^\mu d_i^{-1}
-v_i\hggama_i^\mu \right)\Psi_P.            \eqn{j^mu2}
\eea
Note that the wave function $\Psi_P$ used here satisfies \eq{Gpole}, and thus
fulfils the normalisation condition (for distinguishable particles) 
\be
i \bar{\Psi}_P
\frac{\partial}{\partial P^2}\left(G_0^{-1}-V\right)\Psi_P = 1.
\eqn{wave_norm}
\ee
This result follows upon taking residues of the identity $G=GG^{-1}G$ and using
the fact that $G^{-1}=G_0^{-1}-V$.
By exposing the bound-state poles in the $G^\mu$ of \eq{G7pt}, one finds
that $j^\mu$ is also the matrix element of the current operator
\be
j^\mu = \la J^\mu \ra \equiv \la K|J^\mu(0)|P\ra    \eqn{j^mu_exp}
\ee
where $K^2 = P^2 = M^2$. In terms of $j^\mu$, the familiar charge and
magnetic form factors of spin-1/2 bound three-body systems are given by
\bea
F_C^2(q^2) &=& \sum |j^0|^2    \eqn{F_C} \\
F_M^2(q^2) &=& \frac{2M_N^2}{\mu^2Q^2}\sum 
\left[-\frac{Q^2}{4M^2}|j^0|^2+\left(1+\frac{Q^2}{4M^2}\right)|{\bf j}|^2\right]
                                     \eqn{F_M}
\eea
where $\sum$ indicates a sum over final spins and an average over initial spins
(recall that states $|P\ra$ and wave functions $\Psi_P$ are to be understood as
having implicit spin and isospin labels). In \eq{F_M} $\mu$ is the magnetic
moment of the bound state, $M_N$ is the nucleon mass, and $Q^2=-q^2$.

\subsubsection{Role of subtraction term in the bound-state current}

When used to calculate the bound-state current $j^\mu$, the role of the
subtraction term in \eq{hello} is seen especially well by writing the
bound-state wave function in terms of its usual Faddeev components:
\be
\Psi = \sum_{i=1}^3 \Psi_i,      \eqn{sum_Psi_i}
\ee
where $\Psi$ is written without a momentum subscript to save on notation, and
where
\be
\Psi_i = D_{0i}v_i\Psi    .   \eqn{Psi_i}
\ee
For this purpose it is sufficient to consider just the one-body current
contributions to $j^\mu$,
\be
j^\mu_{\scriptsize\mbox{one-body}} = 
\sum_{i=1}^3\bar{\Psi}\hggama_i^\mu \left(D_{0i}^{-1}-v_i\right)\Psi. 
\ee 
Using \eq{Psi_i} this then evaluates to
\be
j^\mu_{\scriptsize\mbox{one-body}}
=\sum_{i=1}^3\bar{\Psi}\hggama_i^\mu D_{0i}^{-1}(\Psi_j+\Psi_k).
\eqn{j^mu_one-body}
\ee
This equation is not symmetrical with respect to the initial and final state
wave functions. That this must be so can be seen as follows. The term
$\hggama_i^\mu D_{0i}^{-1}$ corresponds to a first order electromagnetic
interaction with particle $i$. During this interaction, particles $j$ and $k$
can be interacting, and this contribution is contained fully in the final state
wave function $\bar{\Psi}$. To avoid overcounting, there must be no preceding
$j-k$ interaction coming from the initial state wave function. This indeed is
the case since, as is evident from \eq{Psi_i}, the very last interaction in
$\Psi_i$ is between particles $j$ and $k$, and just this component is missing
from the right-hand side (RHS) of \eq{j^mu_one-body}.

Another way to write \eq{j^mu_one-body} is
\be
j^\mu_{\scriptsize\mbox{one-body}} =\bar{\Psi}\Gamma^\mu_0\Psi
-\sum_{i=1}^3\bar{\Psi}\hggama_i^\mu D_{0i}^{-1}\Psi_i \eqn{j^mu_subtract}
\ee
where $\Gamma ^\mu_0=\sum_{i=1}^3 \hggama_i^\mu D_{0i}^{-1}$ is the
electromagnetic vertex function in the impulse approximation. In a
three-dimensional approach, the term $\bar{\Psi}\Gamma^\mu_0\Psi$ would
correspond to the correct expression for the total one-body current. In a
four-dimensional approach, however, one needs to subtract the last term of
\eq{j^mu_subtract} in order to avoid the overcounting just discussed. In this
regard we note that the pioneering four-dimensional calculation of Ref.\
\cite{RT} used $\bar{\Psi}\Gamma^\mu_0\Psi$ as the expression for the one-body
current, and therefore the results of this calculation contain overcounting.
\begin{figure}[b]
\hspace*{1cm}  \epsfxsize=14cm\epsfbox{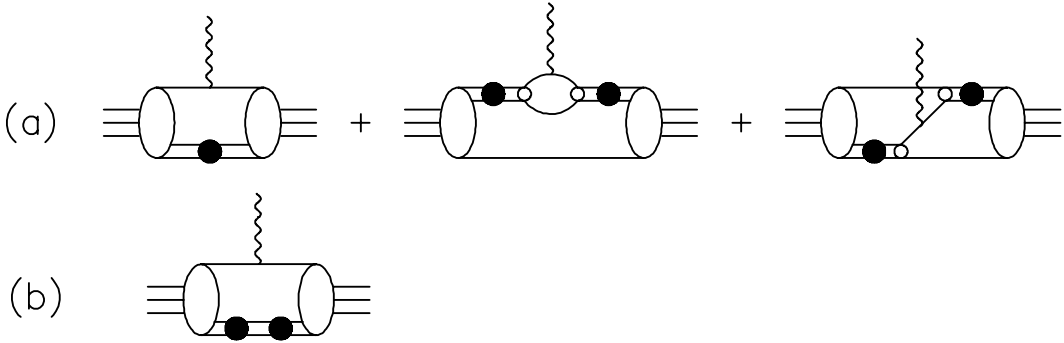}
\vspace{5mm}
\caption{\fign{jmu} (a) The one-body electromagnetic current when two-body 
  interactions are separable. (b) The overcounting term present in
  $\bPsi\Gamma_0^\mu \Psi$. The double-line with the black dot represents the
  dressed quasiparticle propagator $\tau$, the small open circle represents the
  separable potential form factor $h$, and the large ellipses represent the
  spectator bound-state form factor $X$.}
\end{figure}

Further insight into the nature of the overcounting problem can be gained by
assuming, as was done in Ref. \cite{RT}, that the input two-body potentials are
separable.  In this case we have
\be
v_i = h_i \lambda h_i
\ee
where $h_i$ is the separable potential form factor with particle $i$ being a
spectator. The two-body $t$-matrix is then given by
\be
t_i = h_i \tau_i h_i 
\ee
where
\be
\tau_i^{-1} = \lambda^{-1} - h_i D_{0i} h_i.
\ee
For separable potentials we follow Ref.\ \cite{RT} and
introduce the ``spectator bound-state form factor'' $X_i$ defined by
\be
\Psi_i =G_0 h_i \tau_i X_i.
\ee
As the Faddeev equation for the wave function components is
\be
\Psi_i = D_{0i} t_i \left( \Psi_j + \Psi_k \right)
\ee
where $ijk$ are cyclic, it follows that
\be
X_i = d_i^{-1}h_i \left(\Psi_j+\Psi_k\right).
\ee
Using these relations in \eq{j^mu_one-body} it is easy to show that
\be
j^\mu_{\scriptsize\mbox{one-body}} =\sum_i\left( \bar{X}_i d_i^\mu \tau_i X_i
+\sum_{j,k\ne i} \bar{X}_j\tau_j h_j D_{0i} d_i^\mu h_k\tau_k X_k \right)
\eqn{jmuX}
\ee
which has a straightforward graphical interpretation. Depicted in \fig{jmu}(a)
are the topologically distinct contributions to the sums in \eq{jmuX}. One can
similarly write
\be
\bPsi \Gamma_0^\mu \Psi = \sum_i\left( \bar{X}_i d_i^\mu \tau_i X_i
+\sum_{j,k\ne i} \bar{X}_j\tau_j h_j D_{0i} d_i^\mu h_k\tau_k X_k 
+ \bar{X}_i d_i^\mu \tau_i\lambda_i^{-1}\tau_i X_i\right)    \eqn{squirrel}
\ee
which compared to the previous equation, contains an extra term given by
\be
\bPsi \hggama_i^\mu v_i\Psi = \sum_i
\bar{X}_i d_i^\mu \tau_i\lambda_i^{-1}\tau_i X_i .
\ee
This term, depicted in \fig{jmu}(b), represents what is being overcounted in
\eq{squirrel}. Considering this overcounting term together with the first term
in \eq{squirrel} [depicted by the first term in \fig{jmu}(a)], we see that the
role of the overcounting term is to overdress the quasiparticle propagator while
the spectator particle is being gauged. Thus, in the separable case, the use of
the expression $\bPsi \Gamma_0^\mu \Psi$ to calculated the one-body
electromagnetic current is equivalent to overdressing the quasiparticle
propagator in the first term of \fig{jmu}(a).

\subsubsection{Gauge invariance of the bound-state current}

To prove the gauge invariance of \eq{j^mu2} for the three-body bound-state
current, it is convenient to use a symbolic notation for the WT identities.  To
do this we introduce the operators $\he_i$ whose numerical form is defined by
\be 
\he_i(k_1k_2k_3,p_1p_2p_3)=ie_i(2\pi)^{12}\delta^4(k_i-p_i-q)
\delta^4(k_j-p_j)\delta^4(k_k-p_k)
\ee
where $ijk$ represents a cyclic ordering of $123$. Then the WT identities for
the gauged two-body potential and gauged one-particle propagator can be written
in three-particle space in terms of commutators as
\be
q_\mu v^\mu_i I_i=[\he_j+\he_k,v_i],\hspace{1cm}
q_\mu d^\mu_i I_j I_k=[\he_i,d_i] \eqn{WT_comm}
\ee
where $I_i$, $I_j$, and $I_k$ are unit operators in the space of particles $i$,
$j$, and $k$, respectively.

Writing \eq{hello} in the form
\be
\Gamma ^\mu =\sum_{i=1}^3 \left(d^{-1}_id^\mu_id^{-1}_id^{-1}_jd^{-1}_k
+v_i^\mu d^{-1}_i-v_i d^{-1}_id^\mu_id^{-1}_i\right)
\eqn{Gam3}
\ee
and using the WT identities of \eq{WT_comm}, the first term on the RHS of
\eq{Gam3} gives
\be
q_\mu \sum_{i=1}^3 d^{-1}_id^\mu_id^{-1}_id^{-1}_jd^{-1}_k=-[\he,G_0^{-1}]
\ee
where
\be
\he=\he_1+\he_2+\he_3      \eqn{e}.
\ee
Similarly for the last two terms of \eq{Gam3} we have that
\be
q_\mu\left(v^\mu_i d^{-1}_i-v_i d^{-1}_id^\mu_id^{-1}_i\right)
= [\he_j+\he_k,v_i]d^{-1}_i+v_i[\he_i,d_i^{-1}]=[\he,v_id^{-1}_i]
\ee
and therefore
\be
q_\mu\sum_{i=1}^3 \left(v^\mu_i d^{-1}_i-v_i d^{-1}_id^\mu_id^{-1}_i\right)
=[e,V]
\ee
where $V$ is the full potential as given by \eq{Vsum}. We have thus shown that
\be
q_\mu \Gamma ^\mu = -[\he,G^{-1}]  \eqn{Gamma-wti3}
\ee
where $G^{-1}=G_0^{-1}-V$ is the inverse of the full Green function. It is
recognised that \eq{Gamma-wti3} is just the shorthand three-particle version of
the two-particle result given in \eq{Gamma-wti}. Current conservation and
therefore the gauge invariance of the bound-state current follows
immediately. It is worth noting that the presence of the subtraction term [last
term in \eq{Gam3}] is essential for current conservation.

\subsubsection{Generalization to all transition currents}

Electromagnetic currents for all possible transitions can be constructed by
following the same procedure as above for the three-body bound-state
current. That is, one can begin with the expression $G^\mu=G\Gamma^\mu G$ and
then take residues at the relevant initial and final state poles. In this
respect it is important to note that the Green function $G$ has poles
corresponding not only to single particle propagators and three-body bound
states, but also to two-body bound states. For example, it can be shown that in
the vicinity of a two-body bound state of particles 2 and~3
\be
G(k_1k_2k_3,p_1p_2p_3)\sim i\frac{\psi_{K_1}(k_2k_3)d(k_1)}{K_1^2-m^2}
\bPsout_{K_1k_1}(p_1p_2p_3)
\hspace{1cm}\mbox{as}\hspace{1cm} K_1^2\rightarrow m^2 \eqn{1out}
\ee
where $m$ is the bound-state mass of particles 2 and 3, and $K_1=k_2+k_3$.
Similarly
\be
G(k_1k_2k_3,p_1p_2p_3)\sim i\Psin_{P_1p_1}(k_1k_2k_3)
\frac{d(p_1)\bar{\psi}_{P_1}(p_2p_3)}{P_1^2-m^2}
\hspace{1cm}\mbox{as}\hspace{1cm} P_1^2\rightarrow m^2 \eqn{1in}
\ee
where $P_1=p_2+p_3$. The seven-point function $G^\mu$ has the same poles as $G$
and one can show, for example, that in the vicinity of these poles
\be
G^\mu(k_1k_2k_3,p_1p_2p_3)\sim i\frac{\psi_{K_1}(k_2k_3)d(k_1)}{K_1^2-m^2} 
\la\oout,K_1k_1|J^\mu(0)|P_1p_1,\iin\ra i
\frac{d(p_1)\bar{\psi}_{P_1}(p_2p_3)}{P_1^2-m^2}. \eqn{Gmu}
\ee
Analogous expressions for $G$ and $G^\mu$ hold in the vicinity of {\em any}
poles of $G$ corresponding to physical initial and final states.
The in- and out-wave functions appearing in these expressions are
defined generally by
\bea
\lefteqn{(2\pi)^4\delta^4(P-p_1-p_2-p_3)\Psin_j(p_1p_2p_3)=}\hspace{2cm} \nn
&& \int d^4x_1d^4x_2d^4x_3 e^{i(p_1x_1+p_2x_2+p_3x_3)}
\la 0|T\Psi^{(1)}(x_1)\Psi^{(2)}(x_2)\Psi^{(3)}(x_3)|j,\iin\ra,
\eea
\bea
\lefteqn{(2\pi)^4\delta^4(K-k_1-k_2-k_3)\bPsout_i(k_1k_2k_3)=}\hspace{2cm} \nn
&&\int d^4y_1d^4y_2d^4y_3 e^{-i(k_1y_1+k_2y_2+k_3y_3)}\la\oout,
i|T\bar{\Psi}^{(1)}(y_1)\bar{\Psi}^{(2)}(y_2)\bar{\Psi}^{(3)}(y_3)|0\ra,
\eea
where $|j,\iin\ra$ and $|i,\oout\ra$ are the in- and out-states of quantum field
theory (QFT) for the physical initial and final states of total momentum $P$ and
$K$, respectively. Here we follow the usual convention where index $j=1,2,3$
labels a state where particle $j$ is free and the other two particles are bound,
and $j=0$ labels the state where all three particles are free. In case of an
initial three-body bound state, the index $j$ and label ``in'' are dropped (for
three-body bound states there is no difference between in- and out-states as
only one physical particle is involved). Similar conventions hold for index
$i$. For a given model specified by the three-body potential $V$, the in- and
out-wave functions are specified by the equations
\be
\bPsout_i=\bar{\Psi}_i^{(0)}+\bPsout_i(V-V_i)G_i \hspace{1cm} 
\Psin_j=\Psi^{(0)}_j+G_j(V-V_j)\Psin_i      \eqn{inout}
\ee
where for $i=0$, $V_i=0$, $\Psi_i^{(0)}$ is the wave function of three free
particles, and $G_i$ is the free Green function, while for $i=1,2,3$, $V_i$ is
given by \eq{V_i_short}, $\Psi_i^{(0)}$ is the wave function of a free particle
$i$ and a bound $(jk)$ pair, and $G_i$ is the disconnected Green function with
particles $j$ and $k$ interacting and $i$ being a spectator, i.e.
\be
G_i = d_i D_i.     \eqn{G_i_def}
\ee
\eqs{inout} hold also for three-body bound states if we drop the inhomogeneous
terms $\bPsi_i^{(0)}$ and $\Psi_j^{(0)}$ but otherwise use $i,j=0$.

Taking right and left residues of $G^\mu$ at the initial and final state poles,
it is easy to see from the above equations that the general expression for an
electromagnetic transition current is
\be
j^\mu_{ij}=\la\oout,i|J^\mu(0)|j,\iin\ra =\bPsout_i\Gamma ^\mu\Psin_j\eqn{j_ij}.
\ee
\eq{j_ij} is valid no matter whether the initial or final states consist of
three free particles, one free and two bound particles, or the three-body bound
state. Likewise the proof of current conservation for \eq{j_ij} is the same for
all cases: one uses \eq{Gamma-wti3} and the fact that
\be
\bPsout_iG^{-1}=0; \hspace{1cm} G^{-1}\Psin_j=0 \eqn{hom}
\ee
to deduce
\bea
\lefteqn{q_\mu\bPsout_i\Gamma^\mu\Psin_j=
\int 
\frac{d^4k_1}{(2\pi)^4}\frac{d^4k_2}{(2\pi)^4}
\frac{d^4p_1}{(2\pi)^4}\frac{d^4p_2}{(2\pi)^4}}
\hspace{1cm}\nn
&&i\sum_{l=1}^3 \bPsout_i(k_1k_2k_3)
\left[G^{-1}(k_1k_2k_3,p_l+q)e_l
-e_lG^{-1}(k_l-q;p_1p_2p_3)\right]\Psin_j(p_1p_2p_3)=0.
\eea

\subsection{Gauging the AGS equations}

In the previous subsection we constructed the electromagnetic currents for all
possible transitions of the three-particle system. The expression derived,
\eq{j_ij}, expresses the transition currents in terms of the vertex function
$\Gamma^\mu$ which in turn is given in terms of the gauged three-body potential
$V^\mu$ - see \eq{Gamma^mu3}. Although this formally solves the problem of how
to gauge a three-body system, it needs to be recognised that \eq{j_ij} may not
always be very useful for practical calculations. For example, in the present
four-dimensional case where the three-particle potential is assumed to be of the
simple form given by \eq{Vsum} and \eq{V_i_short}, the numerical evaluation of
\eqs{inout} for the scattering in- and out-wave functions is problematic. This
difficulty is due to the disconnectedness of the three-particle potential which
forms the kernel for these integral equations. For other cases the use of
\eq{j_ij} can similarly be impractical. In the four-dimensional description of
the \piNN\ system \cite{KB4d,PA4d}, the possibility of pion absorption together
with overcounting problems makes the \piNN\ potential $V$ difficult to specify,
let alone calculate. Likewise in the spectator approach to the three-nucleon
problem where spectator nucleons are put on mass shell \cite{Gross}, the
effective three-nucleon potential is not easily revealed.

All these problems can be resolved by avoiding the use of potentials in the
formulation of the strong interaction three-body problem. This is just what is
done in the case of three-dimensional Quantum Mechanics by using the
Alt-Grassberger-Sandhas (AGS) equations \cite{AGS} to describe the strong
interactions of three particles. The AGS equations take as input two-body $t$
matrices, rather than potentials, result (after one iteration) in a connected
kernel, and form one of the standard tools for doing practical three-body
calculations. These same benefits can be obtained in four dimensions by using
equations that are the four-dimensional analogue of the AGS equations. Indeed
such equations (which we also refer to as the AGS equations) have already been
used in the successful formulation of the \piNN\ problem \cite{KB4d,PA4d}, and
in the three-nucleon problem within the spectator approach \cite{Gross}. Thus
for practical reasons it is important to apply our gauging of equations method
directly to the AGS equations themselves. In this way we shall obtain a gauge
invariant generalisation of the AGS formulation to systems consisting of three
particles with an added external photon.

\subsubsection{$j(ki)\rightarrow i(jk)$ transition current}

Our starting point is the four-dimensional version of the AGS equations
describing the scattering of three strongly interacting particles. These AGS
equations can be written in the two forms
\be
U_{ij}=G_0^{-1}\bar{\delta}_{ij}+\sum_{k=1}^{3}\bar{\delta}_{ik}T_k G_0U_{kj}
\hspace{5mm}; \hspace{5mm} U_{ij}=G_0^{-1}\bar{\delta}_{ij}+
\sum_{k=1}^{3}U_{ik}G_0T_k\bar{\delta}_{kj} \eqn{AGS_eq}
\ee
where the AGS amplitude $U_{ij}$ describes the process $j(ki)\rightarrow i(jk)$,
i.e. the scattering of particle $j$ off the $(ki)$ quasi-particle, leading to a
final state consisting of particle $i$ and the $(jk)$ quasi-particle. 
If the quasi-particles form bound states then the amplitude $U_{ij}$ is
related to the $t$-matrix for the physical $j(ki)\rightarrow i(jk)$ process by
\be
T_{ij} = \bar{\psi}_i U_{ij} \psi_j         \eqn{T_ij}
\ee
where $\psi_i$ is the two-body bound-state wave function of the ($jk$) system.
In \eq{AGS_eq}, $T_k$ is defined by
\be
T_k = t_k d_k^{-1}    \eqn{T_k}
\ee
where $t_k$ is the two-particle $t$-matrix for the scattering of particles $i$
and $j$. The AGS equations of \eq{AGS_eq} can be written in matrix form as
\be
\UU=G_0^{-1}{\cal I} + {\cal I} \TT G_0 \UU \hspace{5mm};
\hspace{5mm} \UU=G_0^{-1}{\cal I} + \UU G_0 \TT {\cal I} ,   \eqn{AGS_matrix}
\ee
where the ($i,k$)'th elements of matrices $\UU$, $\TT$, and ${\cal I}$ are
defined by
\bea
[\UU]_{ik} &=& U_{ij}  ,  \\
{[\TT]}_{ik} &=& \delta_{ik}T_k ,       \eqn{Tb_ik} \\
{[{\cal I}]}_{ik}  &=& \bar{\delta}_{ik} = 1-\delta_{ik} . \eqn{I_ik}
\eea

Although one could now gauge the matrix $\UU$ by gauging \eqs{AGS_matrix} in the
usual way, the presence of $G_0^{-1}$ in \eqs{AGS_matrix} makes it more
convenient to instead gauge the Green function quantity
\be
\tUU = G_0 \UU G_0
\ee
which satisfies the equations
\be
\tUU={\cal I}G_0+{\cal I}G_0\TT\tUU
\hspace{5mm}; \hspace{5mm}
\tUU=G_0{\cal I}+\tUU\TT G_0{\cal I}  .    \eqn{tildeU}
\ee
At this stage it is convenient to make use of the two-body bound-state
vertex function $\phi_i$, defined as in \eqs{phi} by 
\be
\psi_i = D_{0i} \phi_i\hspace{5mm};\hspace{5mm} \bar{\psi_i} =
\bar{\phi}_iD_{0i}.
\ee
Then the $t$-matrix of \eq{T_ij} can be written as
\be
T_{ij} = \bar{\phi}_i D_{0i} U_{ij} D_{0j}\phi_j 
= \bar{\phi}_i d_i^{-1} \tU_{ij} d_j^{-1} \phi_j . \eqn{Tij}
\ee
As previously discussed, gauged potentials and $t$-matrices do not have photons
attached to external legs. Yet such attachments are necessary for gauge
invariance. For this reason we do not gauge the $t$-matrix $T_{ij}$, but instead
introduce the quantity
\be
\tilde{T}_{ij}=d_iT_{ij} d_j
\ee
which contains extra propagators for the initial and final state spectator
particles. It is by gauging $\tilde{T}_{ij}$ that photons get attached to all
possible places in the $j(ki)\rightarrow i(jk)$ process. Of course after
gauging, it is necessary to remove these propagators to get the corresponding
electromagnetic transition current
$j_{ij}^\mu$, thus
\be
j_{ij}^\mu = d_i^{-1} \tilde{T}_{ij}^\mu  d_j^{-1} . \eqn{calT_ij}
\ee

From \eq{Tij} we find that
\be
\tilde{T}_{ij} = \bar{\phi}_i \tU_{ij} \phi_j.    \eqn{tT_ij}
\ee
Gauging this equation gives
\be
\tilde{T}^\mu_{ij}=\bar{\phi}^\mu_i\tilde {U}_{ij}\phi_j+
\bar{\phi}_i\tilde {U}_{ij}\phi_j^\mu+
\bar{\phi}_i\tilde {U}^\mu_{ij}\phi_j .      \eqn{T^mu_ij}
\ee
Here $\bar{\phi}^\mu_i$ and $\phi_j^\mu$ are the gauged two-body bound-state
vertex functions discussed in Sec.\ II. Being gauged quantities of the two-body
problem, they form an input to the gauged three-body problem. As $\tU_{ij}$
is assumed to be known from the solution of the strong interaction three-body
problem, only the gauged AGS Green function $\tU^\mu_{ij}$ is left to be
determined.  We could find $\tU^\mu_{ij}$ by gauging \eqs{tildeU} explicitly;
however, this is not really necessary as there is a one-to-one correspondence
with the previous gauging of \eq{G3}. Indeed \eq{tildeU} follows formally from
\eq{G3} upon the following substitutions:
\be
G\rightarrow \tUU;\hspace{15mm} G_0\rightarrow{\cal I}G_0;
\hspace{15mm} V\rightarrow \TT.
\ee
Moreover, just as \eq{Vsum} expresses $V$ as a sum of three components
$V_i=v_i d_i^{-1}$, we can similarly write $\TT$ as a
sum
\be
\TT=\WW_1+\WW_2+\WW_3
\ee
where
\be
\WW_i = w_i d_i^{-1}.
\ee
As matrix $\TT$ is specified by \eqs{T_k} and (\ref{Tb_ik}), it follows that
$w_i$ is a matrix whose $(n,m)$'th element is given by
\be
[w_i]_{nm}=\delta_{ni}\delta_{im}t_i.
\ee
With the correspondence now complete, we can immediately use \eq{hello}
to write down the gauged matrix $\tUU^\mu$. We obtain
\be
\tUU^\mu=\tUU\Gamma ^\mu\tUU     \eqn{tU^mu}
\ee
where $\Gamma^\mu$ is a matrix whose $(n,m)$'th element is
\bea
\Gamma^\mu_{nm} &=& 
\sum_{i=1}^3 \left({\cal I}^{-1}_{nm} \hggama_i^\mu D_{0i}^{-1}
+\delta_{ni}\delta_{im}t_i^\mu  d_i^{-1}
-\delta_{ni}\delta_{im} t_i \hggama_i^\mu\right)  \\
&=& {\cal I}^{-1}_{nm} \sum_{i=1}^3  \hggama_i^\mu D_{0i}^{-1}
+ \delta_{nm} t_n^\mu d_n^{-1} - \delta_{nm} t_n \delta_n^\mu . \eqn{Umu}
\eea
Alternatively, we may write $\Gamma^\mu$ in matrix form as
\be
\Gamma^\mu = \sum_{i=1}^3 \left({\cal I}^{-1} \hggama_i^\mu D_{0i}^{-1}
+ w_i^\mu d_i^{-1} - w_i \hggama_i^\mu\right)   . \eqn{Gamma^mu_matrix}
\ee
The use of the same symbol $\Gamma^\mu$ for both the vertex function of
\eq{hello} and the matrix of \eq{Gamma^mu_matrix} should not cause confusion as
only the latter appears in matrix expressions.  On the other hand, using the
same symbol has the advantage of emphasising the formal similarity between the
two. For example, the matrix form of \eq{Gamma^mu_matrix} is equally well
illustrated by \fig{gamma_3d} with $v$'s replaced by $w$'s.

\subsubsection{Current conservation }

In this subsection we would like to show that the current $j^{\mu}_{ij}$ as
specified by \eqs{calT_ij} and (\ref{T^mu_ij}) is conserved.  To do this we
first show that the gauged AGS Green function $\tUU^\mu$ satisfies the
Ward-Takahashi identity. Writing \eq{Gamma^mu_matrix} as
\be
\Gamma^\mu = {\cal I}^{-1} G_0^{-1}G_0^\mu G_0^{-1}
+ \sum_{i=1}^3 \left(w_i^\mu d_i^{-1} - w_i \hggama_i^\mu\right), \eqn{fruit}
\ee
we may use the Ward-Takahashi identities for the input quantities $w_i^\mu$
and $\ggama_i^\mu$:
\bea
q_\mu w_i^\mu (k_jk_k;p_jp_k)&=&i\sum_{l\neq i}[e_lw_i(k_l-q;p_jp_k)-
w_i(k_jk_k;p_l+q)e_l]\\
q_\mu \ggama_i^\mu (k_i,p_i)&=&i\left[d_i^{-1}(k_i)e_i- e_id_i^{-1}(p_i)\right],
\eea
where the former equation follows from identical arguments to that proving
\eq{qVmu}, and where the latter equation follows from
\be
q_\mu d_i^\mu (k_i,p_i)=i[e_id_i(p_i)- d_i(k_i)e_i] .
\ee
From the Ward-Takahashi identity for $G_0^\mu$, it is also easy to show that
\be
q_\mu \Gamma_0^\mu(k_1k_2k_3;p_1p_2p_3) = i\sum_{i=1}^3
\left[G^{-1}_0(k_1k_2k_3;p_i+q)e_i - e_iG^{-1}_0(k_i-q;p_1p_2p_3)\right]
\ee
where
\be
\Gamma_0^\mu = G_0^{-1}G_0^\mu G_0^{-1}.
\ee
Using these results in \eq{fruit} gives
\bea 
\lefteqn{q_\mu \Gamma^\mu (k_1k_2k_3;p_1p_2p_3)=
i\sum_{i=1}^3 \left\{ {\cal I}^{-1}[G^{-1}_0(k_1k_2k_3;p_i+q)e_i -
e_iG^{-1}_0(k_i-q;p_1p_2p_3)] \right. }\hspace{3cm} \nn && +
d_i^{-1}(k_i,p_i)\sum_{l\neq i}[e_lw_i(k_l-q;p_jp_k)-w_i(k_jk_k;p_l+q)e_l]
\nn && \,\left.
-\hw_i(k_jk_k;p_jp_k)[d_i^{-1}(k_i)e_i- e_id_i^{-1}(p_i)]\right\}
\eqn{hithere}
\eea
As $\sum_{i=1}^3\sum_{l\ne i} a_ib_l = \sum_{i=1}^3\sum_{l\ne i} a_lb_i$, we
can rewrite \eq{hithere} as
\bea 
\lefteqn{q_\mu \Gamma^\mu (k_1k_2k_3;p_1p_2p_3)=} \nn
&& i\sum_{i=1}^3 \left[ {\cal I}^{-1}G^{-1}_0(k_1k_2k_3;p_i+q)
- \sum_{l\neq i}   d_l^{-1}(k_l,p_l)w_l(k_mk_n;p_i+q)
-\hw_i(k_jk_k;p_jp_k)d_i^{-1}(k_i) \right] e_i \nn
&&\hspace*{-7mm}
-i\sum_{i=1}^3 e_i \left[ {\cal I}^{-1}G^{-1}_0(k_i-q;p_1p_2p_3)
- \sum_{l\neq i} d_l^{-1}(k_l,p_l)w_l(k_i-q;p_mp_n)
-\hw_i(k_jk_k;p_jp_k)d_i^{-1}(p_i) \right] \nn
&&                    \eqn{big1}
\eea
where $(ijk)$ and $(lmn)$ are both cyclic permutations of $(123)$. Recognising
that the square bracket terms of \eq{big1} correspond exactly to the expression
for $\tUU^{-1}$ derived from \eqs{tildeU}, we may write
\be
q_\mu\Gamma^\mu(k_1k_2k_3;p_1p_2p_3) =
i\sum_{i=1}^3 \left[ \tUU^{-1}(k_1k_2k_3;p_i+q)e_i
-e_i\tUU^{-1}(k_i-q;p_1p_2p_3) \right] .
\ee
Using \eq{tU^mu}, we thus obtain the Ward-Takahashi identity for $\tUU^\mu$:
\be
q_\mu\tUU^\mu(k_1k_2k_3;p_1p_2p_3)=i\sum_{i=1}^3
\left[ e_i\tUU(k_i-q;p_1p_2p_3)-\tUU(k_1k_2k_3;p_i+q)e_i\right] .
\eqn{WTI_U}
\ee

Next we use this result and the last term of \eq{T^mu_ij} to write
\bea
\lefteqn{q_\mu\bar{\phi}_i\tilde {U}^\mu_{ij}\phi_j(k_iK_i;p_jP_j) =
i\int (2\pi)^{-8}\delta^4(K_i-k_j-k_k)\delta^4(P_j-p_i-p_k)
d^4k_j\, d^4k_k\, d^4p_k\, d^4p_i}\hspace*{1cm}\nn
&& \left\{\bar{\phi}_i(k_jk_k;K_i)\left[ e_i\tU_{ij}(k_i-q;p_1p_2p_3)-
\tU_{ij}(k_1k_2k_3;p_j+q)e_j\right] \phi_j(p_kp_i;P_j)\right.\nn
&& \hspace*{2cm} 
+\sum_{l\neq i}\bar{\phi}_i(k_jk_k;K_i)e_l\tU_{ij}(k_l-q;p_1p_2p_3)
\phi_j(p_kp_i;P_j) \nn
&& \hspace*{2cm}
- \sum_{l\neq j} \left. \bar{\phi}_i(k_jk_k;K_i)
\tU_{ij}(k_1k_2k_3;p_l+q)e_l\phi_j(p_kp_i;P_j) \right\}  \eqn{qU}
\eea
where the momentum variables are as specified in \fig{U}.
\begin{figure}[t]
\hspace*{6cm}  \epsfxsize=4.0cm\epsfbox{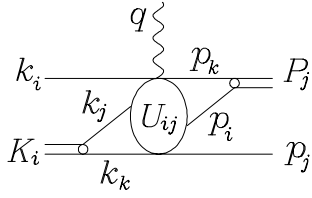}
\vspace{3mm}
\caption{\fign{U} Illustration of the variables involved in the evaluation of
$q_\mu\bar{\phi}_i\tilde {U}^\mu_{ij}\phi_j(k_iK_i;p_jP_j)$ as in \protect
\eq{qU}.}
\end{figure}
To find the contribution of this term to $q_\mu j^\mu_{ij}$, we need to
multiply it by $d_i^{-1}(k_i)d_j^{-1}(p_j)$ and then take the on mass shell
limit, i.e. $d_i^{-1}(k_i)\rightarrow 0$ and $d_j^{-1}(p_j) \rightarrow
0$. Doing this we see that the first term in the curly brackets will give zero
and in this sense is gauge invariant. On the other hand, the last two terms in
the curly brackets will not give zero since the factor
$d_i^{-1}(k_i)d_j^{-1}(p_j)$ will be cancelled by propagators $d_i(k_i)$ and
$d_j(p_j)$ contained in the $\tU_{ij}$. Thus, the contribution of the last two
terms of \eq{qU} to $j^\mu_{ij}$ will not be gauge invariant. However, it
is now easy to check that gauge invariance is restored by including the other
two terms defining $\tT_{ij}^\mu$ in \eq{T^mu_ij}. To do this, we change
integration variables in the last two terms of \eq{qU}:
\bea
\lefteqn{q_\mu\bar{\phi}_i\tilde {U}^\mu_{ij}\phi_j(k_iK_i;p_jP_j) =
i\int d^4k_j\, d^4k_k\, d^4p_k\, d^4p_i \, (2\pi)^{-8}
 \left\{  \delta^4(K_i-k_j-k_k)\delta^4(P_j-p_i-p_k) \right. }\nn
&& \bar{\phi}_i(k_jk_k;K_i)\left[ e_i\tU_{ij}(k_i-q;p_1p_2p_3)-
\tU_{ij}(k_1k_2k_3;p_j+q)e_j\right] \phi_j(p_kp_i;P_j) \nn
&& +\delta^4(K_i-k_j-k_k-q)\delta^4(P_j-p_i-p_k)
\sum_{l\neq i}\bar{\phi}_i(k_l+q;K_i)e_l
\tU_{ij}(k_1k_2k_3;p_1p_2p_3)\phi_j(p_kp_i;P_j) \nn
&&- \delta^4(K_i-k_j-k_k)\delta^4(P_j+q-p_i-p_k)
\sum_{l\neq j} \left. \bar{\phi}_i(k_jk_k;K_i)
\tU_{ij}(k_1k_2k_3;p_1p_2p_3)e_l\phi_j(p_l-q;P_j) \right\} .\nn
&&       \eqn{qUU}
\eea
Making use of the WT identity of \eq{phi-wti}, we have that
\be
q_\mu\phi_j^\mu(p_kp_i;P_j) = i\sum_{l\ne j}e_l\phi_j(p_l-q;P_j) , \hspace{1cm}
q_\mu\bphi_i^\mu(k_jk_k;K_i)=-i\sum_{l\ne i} \bphi_i(k_l+q;K_i)e_l ,
\eqn{phi-bphi-wti}
\ee
and it becomes evident that the last two terms of \eq{qUU} correspond exactly
to $-q_\mu\bphi_i^\mu \tU_{ij}\phi_j-\bphi_i\tU_{ij}q_\mu\phi^\mu_j$.
Thus contracting \eq{T^mu_ij} with $q_\mu$ gives
\bea
\lefteqn{q_\mu\tilde{T}^\mu_{ij}=i \int d^4k_j\, d^4k_k\, d^4p_k\, d^4p_i\,
(2\pi)^{-8}  \delta^4(K_i-k_j-k_k)\delta^4(P_j-p_i-p_k)} \nn
&&\bar{\phi}_i(k_jk_k;K_i)\left[e_i\tU_{ij}(k_i-q;p_1p_2p_3)-
\tU_{ij}(k_1k_2k_3;p_j+q)e_j\right] \phi_j(p_kp_i;P_j) ,
\eea
and the current conservation of $j^\mu_{ij}$ follows.

\subsubsection{$\tUU^\mu$written without subtraction terms}

The gauged AGS Green function $\tU^\mu_{nm}$ is expressed by \eqs{tU^mu} and
(\ref{Umu}) as
\be
\tU^\mu_{nm} = \sum_{i=1}^3 \left[ \left(\tUU {\cal I}^{-1}\hggama^\mu_i
D_{0i}^{-1}\tUU\right)_{nm} + \tU_{ni} t_i^\mu d_i^{-1} \tU_{im}
-  \tU_{ni} t_i \hggama_i^\mu \tU_{im}    \right].    \eqn{U^mu_nm}
\ee
This form for $\tU^\mu_{nm}$ may be the best for practical calculations, but the
presence of the minus sign in the subtraction term $- \tU_{ni} t_i \hggama_i^\mu
\tU_{im}$ does make the perturbation theory expansion of $\tU^\mu_{nm}$
difficult to see. Indeed, if we were to use \eq{U^mu_nm} directly for this
purpose, we would need to carefully keep track of the cancellations between
contributions of the subtraction term and the contributions coming from the term
$\left(\tUU {\cal I}^{-1}\hggama^\mu_i D_{0i}^{-1}\tUU\right)_{nm}$. For this
reason we would like to find an alternative expression for $\tU^\mu_{nm}$
where  all terms contribute with a positive sign.

In order to expose the term in $\left(\tUU {\cal I}^{-1}\hggama^\mu_i
D_{0i}^{-1}\tUU\right)_{nm}$ that will cancel the subtraction term, we use
the AGS equations, \eqs{tildeU}, to write
\bea
\lefteqn{\left(\tUU {\cal I}^{-1}\hggama^\mu_i
D_{0i}^{-1}\tUU\right)_{nm} = 
\left[(1+\tUU T) d^\mu_i D_{0i}\,{\cal I}(T\tUU+1]\right]_{nm}=
\bar{\delta}_{nm} d^\mu_i D_{0i} }\nn    
&&+\sum_{l=1}^3\tU_{nl}t_l d^{-1}_l\bar{\delta}_{lm} d^\mu_iD_{0i}+    
\sum_{k=1}^3 d^\mu_iD_{0i}\bar{\delta}_{nk}t_k d^{-1}_k\tU_{km}+    
\sum_{l,k=1}^3\tU_{nl}t_l d^{-1}_l\bar{\delta}_{lk} d^\mu_iD_{0i}t_k
d^{-1}_k \tU_{km}.\hspace{5mm}
\eea
The last term in this equation can be written as
\ben
\sum_{l,k=1}^3\tU_{nl}t_l d^{-1}_l\bar{\delta}_{lk} d^\mu_iD_{0i}t_k
d^{-1}_k \tU_{km}=
\sum_{l,k=1}^3\tU_{nl}t_l d^{-1}_l(\delta_{li}+\bar{\delta}_{li})
\bar{\delta}_{lk}
d^\mu_iD_{0i}(\delta_{ik}+\bar{\delta}_{ik})t_k d^{-1}_k\tU_{km}.
\een
Expanding out the brackets we obtain four terms, three of which can be
simplified:
\bea
\sum_{l,k=1}^3\tU_{nl}t_l d^{-1}_l\delta_{li}
\bar{\delta}_{lk}
d^\mu_iD_{0i}\delta_{ik}t_k d^{-1}_k\tU_{km} &=& 0 , \\
\sum_{l,k=1}^3\tU_{nl}t_l d^{-1}_l\delta_{li}
\bar{\delta}_{lk}
d^\mu_iD_{0i}\bar{\delta}_{ik}t_k d^{-1}_k\tU_{km} &=&
\tU_{ni}t_i d^{-1}_i d^\mu_i d^{-1}_i
\left(\tU_{im}-G_0\bar{\delta}_{im}\right),
\\
\sum_{l,k=1}^3\tU_{nl}t_l d^{-1}_l\bar{\delta}_{li} \bar{\delta}_{lk}
d^\mu_iD_{0i}\delta_{ik}t_k d^{-1}_k\tU_{km} &=&
\left(\tU_{ni}-G_0\bar{\delta}_{ni}\right)
d^{-1}_i d^\mu_i d^{-1}_it_i\tU_{im} .
\eea
Thus
\bea
\left(\tUU {\cal I}^{-1}\hggama^\mu_i
D_{0i}^{-1}\tUU\right)_{nm} &=&
\bar{\delta}_{nm} d^\mu_iD_{0i}+
\sum_{l=1}^3\tU_{nl}t_l d^{-1}_l\bar{\delta}_{lm} d^\mu_iD_{0i}+    
\sum_{k=1}^3 d^\mu_iD_{0i}\bar{\delta}_{nk}t_k d^{-1}_k\tU_{km}  \nn
&+&
\tU_{ni}t_i d^{-1}_i d^\mu_i d^{-1}_i
\left(\tU_{im}-G_0\bar{\delta}_{im}\right)+
\left(\tU_{ni}-G_0\bar{\delta}_{ni}\right)
d^{-1}_i d^\mu_i d^{-1}_it_i\tU_{im}\nn
&+&\sum_{l,k=1}^3\tU_{nl}t_l d^{-1}_l\bar{\delta}_{li} \bar{\delta}_{lk}
d^\mu_iD_{0i}\bar{\delta}_{ik}t_k d^{-1}_k\tU_{km} .
\eea
Using that
\bea  
\lefteqn{\sum_{l=1}^3\tU_{nl}t_l d^{-1}_l\bar{\delta}_{lm} d^\mu_iD_{0i}
-\tU_{ni}t_i d^{-1}_i d^\mu_i d^{-1}_iG_0\bar{\delta}_{im} }\nn
&&=\sum_{l=1}^3\tU_{nl}t_l d^{-1}_l\bar{\delta}_{lm} d^\mu_iD_{0i}
[1-\delta_{il}]=\sum_{l=1}^3\tU_{nl}t_l d^{-1}_l
\bar{\delta}_{lm}\bar{\delta}_{li} d^\mu_iD_{0i}
\eea
and similarly 
\bea
\lefteqn{\sum_{k=1}^3 d^\mu_iD_{0i}\bar{\delta}_{nk}t_k d^{-1}_k\tU_{km}
-G_0\bar{\delta}_{ni} d^{-1}_i d^\mu_i d^{-1}_it_i\tU_{im} } \nn
&&=\sum_{k=1}^3 d^\mu_iD_{0i}\bar{\delta}_{nk}t_k d^{-1}_k\tU_{km}
[1-\delta_{ik}]=\sum_{k=1}^3 d^\mu_iD_{0i}
\bar{\delta}_{nk}\bar{\delta}_{ik}t_k d^{-1}_k\tU_{km},
\eea
we obtain that
\bea
\lefteqn{ \left(\tUU {\cal I}^{-1}\hggama^\mu_i
D_{0i}^{-1}\tUU\right)_{nm} =
\bar{\delta}_{nm} d^\mu_iD_{0i}+
\sum_{l=1}^3\tU_{nl}t_l d^{-1}_l\bar{\delta}_{lm}\bar{\delta}_{li}
d^\mu_iD_{0i}+    
\sum_{k=1}^3 d^\mu_iD_{0i}\bar{\delta}_{nk}\bar{\delta}_{ik}t_k d^{-1}_k
\tU_{km} } \hspace{3cm}\nn
&&+2\tU_{ni}t_i d^{-1}_i d^\mu_i d^{-1}_i\tU_{im}
+\sum_{l,k=1}^3\tU_{nl}t_l d^{-1}_l\bar{\delta}_{li}
\bar{\delta}_{lk}
d^\mu_iD_{0i}\bar{\delta}_{ik}t_k d^{-1}_k\tU_{km}.\hspace{5mm}
\eea
Substituting into \eq{U^mu_nm} we finally obtain
\bea
\lefteqn{\tU^\mu_{nm}=\sum_{i=1}^3\left[ \bar{\delta}_{nm} d^\mu_iD_{0i}+
\sum_{l=1}^3\tU_{nl}t_l d^{-1}_l
\bar{\delta}_{lm}\bar{\delta}_{li} d^\mu_iD_{0i}+    
\sum_{k=1}^3 d^\mu_iD_{0i}\bar{\delta}_{nk}\bar{\delta}_{ik}t_k d^{-1}_k
\tU_{km} \right. } \hspace{1cm}\nn 
&&+\left.\tU_{ni}\left(t_i\hggama^\mu_i
+t_i^\mu d^{-1}_i\right)\tU_{im}
+\sum_{l,k=1}^3\tU_{nl}t_l d^{-1}_l\bar{\delta}_{li}
\bar{\delta}_{lk}
d^\mu_iD_{0i}\bar{\delta}_{ik}t_k d^{-1}_k\tU_{km}\right]. \eqn{+}
\eea
\begin{figure}[t]
\hspace{5mm}\epsfxsize=16.0cm\epsfbox{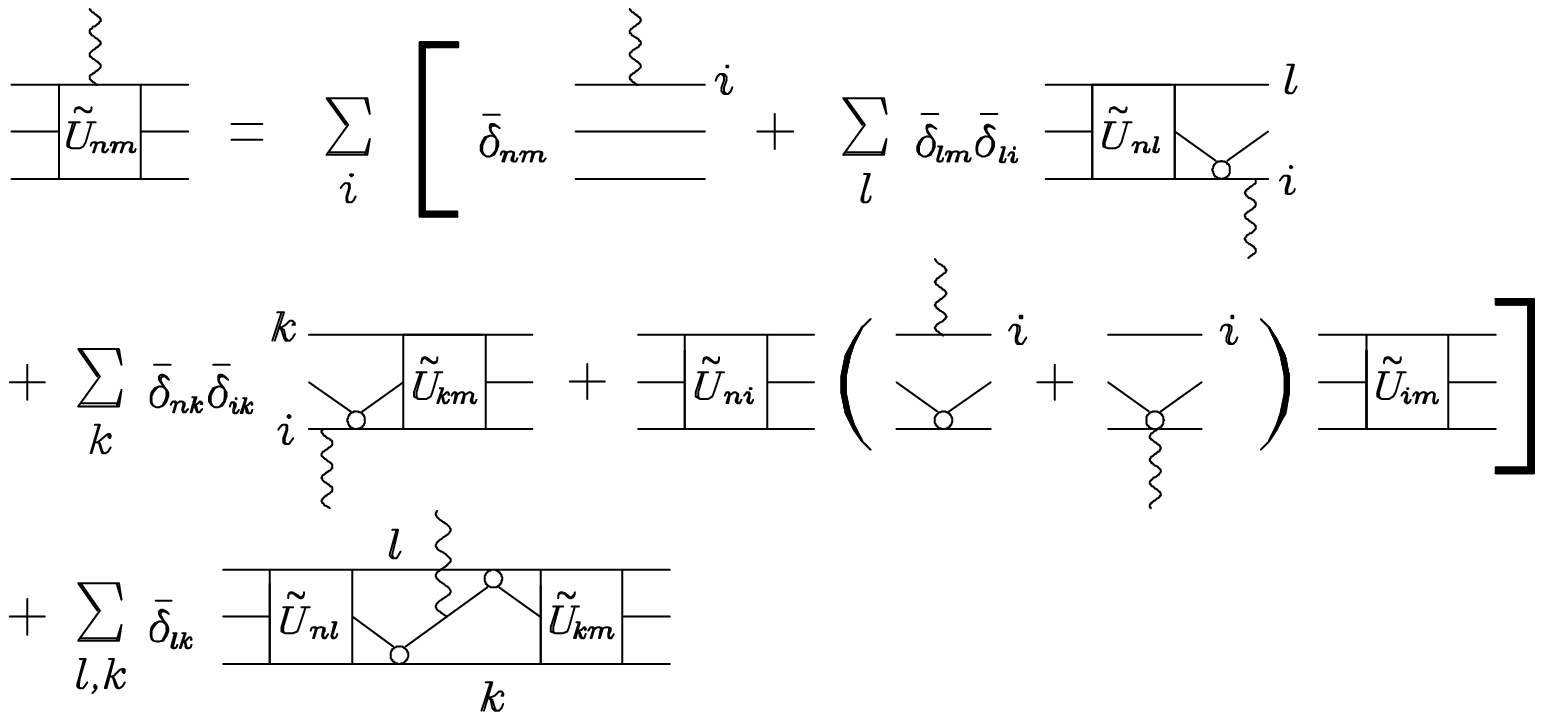}
\caption{\fign{Umunm} Graphical representation of \protect\eq{+} for the
gauged AGS Green function $\tU^\mu_{nm}$.  }
\end{figure}
This form for $\tU^\mu_{nm}$ has no subtraction terms and can be used directly
to generate the corresponding perturbation theory expansion. This is seen
especially well from the graphical representation of \eq{+} given in
\fig{Umunm}. 

Taking left and right residues of $\tUU^\mu$ at the three-body bound-state
poles leads to the three-body bound-state current $j^\mu$ (see Subsec.\ 5
below). In that case only the last three terms of \fig{Umunm} contribute.  If we
then consider the case of one-body currents with separable two-body
interactions, we see that \fig{Umunm} reduces down to \fig{jmu}(a).

\subsubsection{$j(ki)\rightarrow ijk$ transition current }

The strong interaction process $j(ki)\rightarrow ijk$ where the final state
consists of three free particles is described by the $t$-matrix
\be
T_{0j} = U_{0j}\psi_j = U_{0j}D_{0j}\phi_j
\ee
where $U_{0j}$ is given by
\be
U_{0j} = \sum_i T_i G_0 U_{ij} = \frac{1}{2}\sum_i U_{ij}-G_0^{-1}. \eqn{U_0j}
\ee
To find the $j(ki)\rightarrow ijk$ electromagnetic transition current
$j_{0j}^\mu$ (which can be used to describe processes like $pd\rightarrow \gamma
npp$), we proceed as before and define the Green function quantity
\be
\tT_{0j} = G_0 T_{0j} d_j  =
\left(\frac{1}{2}\sum_i \tU_{ij}-G_0\right)\phi_j.  \eqn{tT_0j}
\ee
It is $\tT_{0j}$ which may now be gauged, thereby obtaining
\be
j_{0j}^\mu = G_0^{-1}\tT_{0j}^\mu d_j^{-1}.
\ee
Using the product rule we have that
\be
\tT_{0j}^\mu = \left(\frac{1}{2}\sum_i \tU_{ij}^\mu-G_0^\mu\right)\phi_j
+\left(\frac{1}{2}\sum_i \tU_{ij}-G_0\right)\phi_j^\mu
\ee
and therefore the electromagnetic current can be written as
\be
j_{0j}^\mu = \left(\frac{1}{2}\sum_{ilk} U_{il}G_0\Gamma_{lk}^\mu G_0 U_{kj}
-\Gamma_0^\mu\right)D_{0j}\phi_j + U_{0j}D_{0j}\phi_j^\mu.    \eqn{j_0j^mu}
\ee

\subsubsection{Three-body bound-state current}

The three-body bound-state current was already discussed in Sec.\ III~B
above. There we obtained an expression, \eq{j^mu1}, which gives $j^\mu$ in terms
of the two-body potentials $v_i$ and the gauged two-body potentials
$v_i^\mu$. Here we would like to give an alternative expression that results
from the gauging of the AGS equations. This has the advantage of giving $j^\mu$
in terms of the two-body $t$-matrices $t_i$ and the gauged two-body $t$-matrices
$t_i^\mu$.

We recall that the AGS amplitudes $U_{ij}$ are defined through
the expression for the $3\rightarrow 3$ Green function:
\be
G = G_i\delta_{ij} + G_i U_{ij} G_j    \eqn{U_ij_def}
\ee
where $G_i$ is given by \eq{G_i_def}. 
Thus it is clear that $U_{ij}$ has a pole at the three-body bound state:
\be
U_{ij} \sim i \frac{\chi_i^P \bar{\chi}_j^P}{P^2-M^2}\hspace{1cm}
\mbox{as}\hspace{1cm} P^2\rightarrow M^2
\ee
where
\be
\chi_i^P = G_i^{-1}\Psi_P   .   \eqn{chi_i}
\ee
By writing $U_{ij}=G_0^{-1}\tU_{ij}G_0^{-1}$ we may gauge \eq{U_ij_def} in the
usual way.  Then taking the left and right residues at the three-body bound
state poles and using \eq{tU^mu} we obtain that
\be
j^\mu = \sum_{lk} \bar{\chi}_l^K G_0 \Gamma_{lk}^\mu G_0 \chi_k^P 
\eqn{j^mu_AGS}.
\ee
From \eq{chi_i} it is easy to see that
\be
G_0 \chi_i^P = \Psi_j^P + \Psi_k^P    \eqn{G_0chi_i}
\ee
where $\Psi_j^P$ and $\Psi_k^P$ are Faddeev components of the bound-state wave
function as in \eq{Psi_i}, with $ijk$ defined to be cyclic. By introducing the
column matrix
\renewcommand{\arraystretch}{1.3}
\be
\Psi_P = \left(\begin{array}{c} \Psi_2^P + \Psi_3^P \\
\Psi_3^P + \Psi_1^P \\ \Psi_1^P + \Psi_2^P\end{array}\right) ,
\ee
with the same symbol $\Psi_P$ being used as for the bound-state wave function,
we may write \eq{j^mu_AGS} in the matrix form
\be
j^\mu = \bPsi_K \Gamma^\mu \Psi_P
\ee
giving us a formally identical expression to that of \eq{j^mu1} but where now
each term on the RHS is a matrix. Interpreted as a matrix equation, this result
expresses the current $j^\mu$ in terms of two-body $t$-matrices and gauged
two-body $t$-matrices [see \eq{Gamma^mu_matrix}], while interpreted as a scalar
equation (i.e. not a matrix equation) it expresses $j^\mu$ in terms of two-body
potentials and gauged two-body potentials [see \eq{hello}].

\subsubsection{$(ijk)\rightarrow i(jk)$ transition current}

In the previous subsection we found the $(ijk)\rightarrow (ijk)$ electromagnetic
transition current by first gauging \eq{U_ij_def} for the green function $G$,
and then taking left and right residues at the three-body bound-state poles. By
contrast, in Subsec.\ 1 the $j(ki)\rightarrow i(jk)$ transition current was
found by first taking left and right residues of \eq{U_ij_def} at the two-body
bound-state poles, which leads to \eq{T_ij}, and then gauging this equation.  It
is straightforward to see that the final expressions for the electromagnetic
transition currents do not depend on the order in which the gauging and the
taking of residues is done.

To determine the $(ijk)\rightarrow i(jk)$ electromagnetic transition current it
is convenient to first take the left residue of \eq{U_ij_def} at the two-body
bound-state pole, then gauge the resulting expression, and finally take the
right-hand residue at the three-body bound-state pole. Taking the left
residue of \eq{U_ij_def} at the bound-state pole of particles $j$ and $k$, but
keeping the left propagator for particle $i$, leads to the Green function
quantity
\be
X_{ij} = \bphi_i G_0 U_{ij} G_j = \bphi_i \tU_{ij} G_0^{-1}G_j .
\ee
Gauging this equation, taking the residue at the three-body bound-state pole
on the right, and then eliminating the left propagator of particle $i$, gives
the $(ijk)\rightarrow i(jk)$ electromagnetic transition current:
\be
j_i^\mu = d_i^{-1}\left( \bphi_i^\mu G_0 \chi_i^P + \sum_{lk}\bphi_i G_0
U_{il}G_0\Gamma_{lk}^\mu G_0 \chi_k^P \right).
\ee
This expression can be written using matrix notation as
\be
j_i^\mu = \bar{\Phi}_i^\mu \Psi_P + \bar{\Phi}_i\,\tUU \Gamma^\mu \Psi_P
\ee
where $\bar{\Phi}_i$ and $\bar{\Phi}_i^\mu$ are row matrices whose $j$'th
elements are defined by $d_i^{-1}\bphi_i\delta_{ij}$ and
$d_i^{-1}\bphi_i^\mu\delta_{ij}$ respectively.

\subsubsection{$(ijk)\rightarrow ijk$ transition current}

The $(ijk)\rightarrow ijk$ electromagnetic transition current $j_0^\mu$
describes the photodisintegration of the three-body bound state leading to three
free particles, $\gamma (ijk)\rightarrow ijk$. To find $j_0^\mu$ we may start
with the seven-point function of \eq{G^mu3}, take the right-hand residue at the
three-body bound state, and multiply on the left by $G_0^{-1}$, in this way
obtaining
\be
j_0^\mu = G_0^{-1} G \Gamma^\mu \Psi_P .
\ee
This gives the transition current for the photodisintegration process in terms
of two-body potentials $v_i$ and gauged two-body potentials $v_i^\mu$.

Alternatively, to obtain the corresponding expression in terms of two-body
$t$-matrices $t_i$ and gauged two-body $t$-matrices $t_i^\mu$, we may proceed as
in the previous two subsections and start with \eq{U_ij_def} which may be
written as
\be
G = \delta_{ij}G_j + G_i G_0^{-1} \tU_{ij} G_0^{-1} G_j
= \delta_{ij}G_j + (1+G_0 T_i)\tU_{ij} G_0^{-1} G_j .
\ee
Gauging the latter form of the equation, taking the right-hand residue at the
three-body bound state and then multiplying on the left by $G_0^{-1}$ gives
\be
j_0^\mu = G_0^{-1}\left(G_0^\mu T_i + G_0 T_i^\mu\right) G_0\chi_i^P
+ \sum_{lk} (1+T_i G_0)U_{il}\Gamma_{lk}^\mu G_0\chi_k^P  .
\ee
Now since $(1+T_i G_0)U_{il} = U_{0l}$, the electromagnetic transition current
can also be written as
\be
j_0^\mu = \left(\Gamma_0^\mu G_0 T_i + T_i^\mu\right) G_0\chi_i^P
+ \sum_{lk} U_{0l}\Gamma_{lk}^\mu G_0\chi_k^P
\ee
with $G_0\chi_i^P$ given by \eq{G_0chi_i}.

\section{SUMMARY}

In this article we have presented a general method for incorporating an external
photon into a system of particles whose strong interactions are described
nonperturbatively by integral equations. This method consists of gauging the
integral equations themselves, and has the important feature of coupling the
external photon to all possible places in the strong interaction model. As the
photon is coupled everywhere, gauge invariance of all expressions for
on-mass-shell electromagnetic transition currents is guaranteed.

To discuss the details of our approach we have chosen the case of three
distinguishable particles (with no coupling to two-particle channels) whose
strong interactions are described by standard four-dimensional integral
equations of quantum field theory. This type of three-particle system presents
the simplest case for which no practical gauging method has so far been
available.  In the two-particle sector there have been previous gauging
procedures that established the conserved currents for the \NN\ system \cite{GR}
(no coupling to one-particle channels) and the \piN\ system \cite{Antwerp}
(coupling to the one-nucleon channel included), yet even in these cases the
gauging of equations method provides a much simpler way to derive the same
results.

By gauging the integral equation for the three-body Green function where the
kernel consists of two-body potentials $v_i$, we obtained an expression,
\eq{j_ij}, that describes all possible electromagnetic transition amplitudes of
the three-body system in terms of the $v_i$ and the gauged potentials $v_i^\mu$.
We have also shown how our method can be used to gauge the
Alt-Grassberger-Sandhas equations for three particles in order to get more
practical relations where the electromagnetic transition amplitudes are expressed
in terms of two-body $t$-matrices $t_i$ and gauged $t$-matrices $t_i^\mu$,
see Sec.\ III~C.

Although we have presented the gauging of equations method within the context of
quantum field theory where the integral equations are four-dimensional, it
should be noted that the method itself can be used in a wider context. Indeed we
have already used this method to incorporate an external photon into the
three-dimensional equations of the spectator approach \cite{G3d,nnn3d}.
Similarly, one could apply the gauging procedure to the three-dimensional
approach of time ordered perturbation theory when the equations are expressed in
terms of convolution integrals \cite{KB2}. The gauging method also does not
depend on the nature of the external field involved, so our results remain valid
if the external field is due to a strongly or weakly interacting probe.  In this
sense the gauging of equations method provides the solution to the long standing
problem of how to incorporate an external field into a nonperturbative
description of quarks or hadrons.

\acknowledgments
The authors would like to thank the Australian Research Council for their
financial support.

\end{document}